\documentclass[conference]{IEEEtran}
\IEEEoverridecommandlockouts
\usepackage{cite}
\usepackage{amsmath,amssymb,amsfonts}
\usepackage{algorithm}
\usepackage{algpseudocode}
\usepackage{graphicx}
\usepackage{textcomp}
\usepackage{xcolor}
\usepackage{todonotes}
\usepackage{xspace}
\usepackage{multicol}
\usepackage{booktabs}
\usepackage{adjustbox}
\usepackage[T1]{fontenc}
\usepackage{subfig}
\newcommand{\pred}{\ensuremath{\mathcal{P}}\xspace}
\newcommand{\vidhyatodo}[1]{{\color{blue}~\small{\textsf{[TODO: #1]}}}}
\newcommand{\vidhyatodoorange}[1]{{\color{orange}~\small{\textsf{[INFO: #1]}}}}

\newcommand{\sandeeptodo}[1]{{\color{purple}~\textsf{[TODO: #1]}}}

\newcommand{\br}[1]{\ensuremath{\langle #1 \rangle}}
\newcommand{\closeopen}[1]{\ensuremath{[#1)}\xspace}

\def\BibTeX{{\rm B\kern-.05em{\sc i\kern-.025em b}\kern-.08em
    T\kern-.1667em\lower.7ex\hbox{E}\kern-.125emX}}
\begin{document}

\title{Efficient Two-Layered Monitor for Partially Synchronous Distributed Systems
}

\author{\IEEEauthorblockN{Vidhya Tekken Valapil\IEEEauthorrefmark{1}, Sandeep Kulkarni\IEEEauthorrefmark{2}, Eric Torng\IEEEauthorrefmark{3}, and Gabe Appleton\IEEEauthorrefmark{4}}
\IEEEauthorblockA{Michigan State University\\
East Lansing, MI 48823, USA \\
Email: \IEEEauthorrefmark{1}tekkenva@cse.msu.edu, \IEEEauthorrefmark{2}sandeep@cse.msu.edu, \IEEEauthorrefmark{3}torng@cse.msu.edu, \IEEEauthorrefmark{4}applet14@msu.edu}}

\maketitle
\thispagestyle{plain}
\pagestyle{plain}
\begin{abstract}
Monitoring distributed systems to ensure their correctness is a challenging and expensive but essential problem.
It is challenging because while execution of a distributed system creates a partial order among events, the monitor will typically observe only one serialization of that partial order. This means that even if the observed serialization is consistent with the system specifications, the monitor cannot assume that the system is correct because some other unobserved serialization can be inconsistent with the system specifications. Existing solutions that guarantee identification of all such unobserved violations require some combination of lots of time and large clocks, e.g. $O(n)$ sized Vector Clocks.

We present a new, efficient two-layered monitoring approach that overcomes both the time and space limitations of earlier monitors. The first layer is imprecise but efficient and the second layer is precise but (relatively) inefficient. We show that the combination of these two layers reduces the cost of monitoring by 85-95\%. 
Furthermore, the two-layered monitor permits the use of $O(1)$ sized Hybrid Logical Clocks. 

\end{abstract}


\section{Introduction}
Concurrency among different processes is the key challenge in designing and validating distributed systems because concurrency leads to a partial order (as opposed to a total order) of events in the system. 
In other words, due to concurrency among processes, the exact relative order of all the events in a distributed system is unknown and only the relative order of a subset of events can be declared with confidence.
More specifically, concurrency leads to the possibility of a race condition where some serializations of the partial order are consistent with the system specifications while other serializations of the partial order are inconsistent with the system specifications.
Errors that arise from a race condition can be especially challenging to find if the serializations that violate the system specifications are relatively few in number and thus rarely arise when the system is executed.



One way to deal with this issue is to monitor the execution of the distributed system where the monitor creates a partial order of events that conforms to its observations and then detects whether any serialization of that partial order can produce a violation of the system specifications. 
Ideally, the partial order created by the monitor should be as restrictive as possible so that the set of total orders (serializations) is as small as possible.
For example, the partial order should preserve the causal relationship \cite{logicalClocks} among events.
However, this can still lead to a very large serialization space, especially if communication among processes is relatively infrequent.
The partial order can be tightened significantly when applications rely on some clock synchronization. For example, if applications assume that the clocks of the processes are synchronized to be within $\epsilon$ and the local clock value of an event $f$ is more than $\epsilon$ ahead of the local clock value of an event $e$, then the monitor can rely on a partial order where $e$  occurs before $f$.

With this intuition, we formally define the scope of our paper: developing monitors for partially synchronous distributed systems that detect whether some serialization of events results in a state where a given predicate \pred evaluates to true.
Note that we make the following two assumptions.
We first assume a partially synchronous distributed system where the clocks of processes in the system differ by at most a given parameter $\epsilon$.
We further assume that each process has some variables, and we can represent a system violation as \pred, defined using the variables, such that a violation occurs if \pred evaluates to true.



We face two key challenges in developing efficient monitors. First, the problem of predicate detection (i.e., detecting if the given predicate $\pred$ is true) itself is NP-complete, so unless \mbox{$P\!=\!NP$}, exponential complexity is unavoidable in the worst case. That said, for some predicates, there is the possibility of polynomial time solutions even if \mbox{$P\!\neq\!NP$}. Second, many monitoring solutions require $O(n)$ sized vector clocks to capture the partial order among events \cite{logicalClocks}. 
For partially synchronous systems, there is the potential to reduce the size of the clocks by using Hybrid Vector Clocks \cite{HVC}. However, in the worst case, their size is still $O(n)$. 

In \cite{MonitoringUsingSMT}, we proposed an approach that uses SMT solvers to address the two challenges above. Specifically, it obviates the need for using $O(n)$ sized vector clocks but relies on $O(1)$ sized Hybrid Logical Clocks \cite{HLC}. While circumventing the NP-completeness result is impossible, it uses SMT solvers to benefit from the advances in the state-of-the-art to determine whether a given Boolean formula is satisfiable. 
Towards this end, we construct a Boolean formula that captures constraints that characterize the partial order of events in the system and a violation of a system requirement.
If these constraints are satisfiable, then there exists a serialization of events that leads to a violation of 
the system requirement.






While using SMT solvers and HLC reduces the cost of monitoring, there are scenarios where the time for monitoring is still very high. Especially in the context of \textit{online} monitoring where the goal is to be able to \textit{react} to a violation quickly, we must ensure that the monitor can \textit{keep up} with the application. This will ensure that the lag between the monitor and the application is bounded so that any potential system violations are quickly detected.

We can improve efficiency by sacrificing accuracy, specifically by allowing a monitor to report (1) false negatives (monitor fails to report actual violations), (2) false positives (monitor reports phantom violations),  or (3) both false positives and false negatives.
While sacrificing accuracy is not ideal, all modes of errors may be acceptable in some contexts.
For example, using efficient monitors that suffer from only false negatives means that although violations will remain undetected for a longer time, they should eventually be detected unless the false negative rate is too high or there is something structural about the violation that ensures it will always be a false negative. 
Efficient monitors with only false positives can be useful as a filtering step when used in combination with a less efficient but accurate monitor.
Even efficient monitors that suffer from both false positives and false negatives can potentially be useful as an imperfect filtering step recognizing that some violations may take longer to detect.
In all cases, having an efficient monitor that finds a significant subset of violations and can be deployed is better than an inefficient and accurate monitor that cannot be deployed. For example, consider applications/systems that are required to be lightweight and have time/memory constraints so they cannot add inefficient compute-intensive monitors.

In this paper, we provide a range of monitors that fall in different places on the efficiency and accuracy spectrum.
%
We begin with the efficient monitor that has no false positives but suffers from false negatives from \cite{PredDetHLC_ICDCN}.
We then present a modified monitor that is efficient 
for many predicates encountered in practice
and has no false negatives but suffers from false positives.
%
We then show that our approach can be used to provide a trade-off between false positives and false negatives where the first two monitors are the end points of this trade-off.
Finally, we present a hybrid monitoring approach where one can eliminate both false positives and false negatives by using a two-layered monitoring approach.

\color{black}

\textbf{Contributions of the paper:}
\begin{itemize}
    \item We first show how to modify the efficient HLC-based monitoring algorithm with false negatives from \cite{PredDetHLC_ICDCN} into an efficient monitoring algorithm without false negatives but with potentially many false positives. Even though the extension cannot handle all possible predicates, it handles predicates encountered in practice (conjunctive, arithmetic, violation of mutual exclusion), etc. 
    
    \item 
    We then discuss how the above algorithm can be used in scenarios where one would like to reduce the number of false positives, but can afford some false negatives, by modifying a single parameter $\gamma$ in the algorithm.
    \item We evaluate the effectiveness of the algorithm  with the help of experiments, where we identify the number of false positives/negatives of the monitor when used in detecting conjunctive predicates.
    \item We present a two-layered monitoring algorithm that combines the algorithm that uses HLC with parameter $\gamma$ with a monitoring algorithm from \cite{MonitoringUsingSMT} that uses SMT solvers to perform predicate detection. This two layered monitoring algorithm eliminates all false positives and, depending on $\gamma$, many or all false negatives are eliminated at a reduced cost.
    \item {We evaluate the effectiveness of the two-layered monitoring approach by evaluating the time taken to detect violations of mutual exclusion in an application that uses time division multiplexing.}
\end{itemize}
\textbf{Organization of the paper. } In Section \ref{sec:preliminaries}, we define the system model and the notion of predicate detection, along with a brief discussion of the properties of Hybrid Logical clocks (HLC). In Sections \ref{sec:HLCBasedPredOriginal} and \ref{sec:falsenegativesWithHLC}, we discuss performing predicate detection using HLC and how it can result in false negatives. In Section \ref{sec:eliminateFNwithEpsExt}, we identify how predicate detection using HLC can be extended to eliminate false negatives for predicates encountered in practice. We also discuss how one can use the extension to trade off between false positives and false negatives in Section \ref{sec:reduceFPVwithGamma}. We analyze the effectiveness of the extended approach through experimental analysis in Section \ref{sec:gammaFPVAnalysis}. 
The implications of false positives/false negatives are discussed in Section  \ref{sec:FPVFNVimplications}. In Section  \ref{sec:predDetUsingSMT}, we briefly discuss the monitoring algorithm to perform predicate detection from \cite{MonitoringUsingSMT}. In Section \ref{sec:twolayeredMonitor}, we present a two layered monitoring approach that combines the approaches in Sections \ref{sec:predDetHLCwithgamma} and \ref{sec:predDetUsingSMT} to reduce the computation time associated with monitoring. We evaluate the efficiency of this combined approach in Section \ref{sec:hlcgammaSMTExpeAnalysis}. Finally in Section \ref{sec:relatedWork}, we discuss related work and conclude in Section \ref{sec:conclusion}.

\section{Preliminaries}
\label{sec:preliminaries}
\subsection{System Model}
\label{sec:sysmodel}
We consider a distributed system of $n$ processes where each process $i$ ($0\leq i < n$) is associated with a local physical clock $pt.i$. In this paper we focus on partially synchronous distributed systems, where clocks at the processes are not perfectly synchronized with each other, but have a bounded clock skew $\epsilon$.

Our partially synchronous model is slightly different from that in \cite{consensusWhenInPartialSynch}. 
In \cite{consensusWhenInPartialSynch}, authors consider systems where there are bounds on the relative speed of processors and on message delay. In such systems, they build clocks 
that 
remain within a constant of each other.
In our work, instead of relying on relative speeds of processes, we assume that (by any available means) the clocks of processes are synchronized to be within a bound that is known to the processes. 
(In  Section \ref{sec:gammaFPVAnalysis}, we consider the extension to cases where the clock skew is not precisely known to the monitor.)  
We make no assumptions about message delays.

Each process is associated with a set of variables, and the \textit{state} of a process is identified by the values of its variables.
We categorize the events that can happen at each process into three categories: send events, receive events, and local events. 
A send or a receive event corresponds to the event of a process sending or receiving a message. 
A local event corresponds to an event at a process where the state of the process may change.
The state of a process does not change between any two consecutive events at a process.
Each event $e$ at a process $i$ is also associated with a physical clock value or timestamp $pt.e$, which is the physical clock value of process $i$ when the event $e$ happened.

\subsection{Monitoring and Predicate Detection}
\label{sec:monitorAndPredDet}

We use a centralized monitor that receives relevant information from the processes (details in Section \ref{sec:HLCBasedPredOriginal}). 
We assume that the communication between the processes and the monitor is FIFO and reliable.
While we do not assume any time bounds on this communication, the time (taken for predicate detection) reported is counted from the time the monitor receives the data from the processes. We do not make any assumptions about the reliability of messages among processes. 
Lost messages do not affect the correctness of the monitor. 
This is due to the fact that when a monitor performs predicate detection within a time window, it deals with messages that were sent in that window but were not received within the window. 
We do assume that spurious messages are not received by any process; spurious messages would cause the monitor to behave incorrectly. We do not consider process faults such as crash of processes or monitor. 

We focus on the trade off between monitor efficiency and monitor accuracy 
when performing \textit{predicate detection}. We now briefly define predicate detection in distributed systems in terms of events and variables. 

An event $e$ \textbf{happened before} an event $f$, denoted as $e\rightarrow f$, if and only if one of the following is true,
\begin{itemize}
    \item events $e$ and $f$ happened at the same process and $f$ happened at the process after the event $e$,
    \item  $e$ is a message send event and $f$ is the corresponding message receive event,
    \item $pt.e + \epsilon < pt.f$
    \item there exists an event $g$ such that $e \rightarrow g$ and $g \rightarrow f$,
\end{itemize}


Two events $e$, $f$ are \textbf{concurrent events}, denoted as $e||f$, iff $(e\not\rightarrow f) \wedge (f\not\rightarrow e)$. 

{We note that the above definition is an extension of the happened-before relation by Lamport \cite{logicalClocks} for partially synchronous systems where the clocks of any two processes differ by at most $\epsilon$.}


A \textbf{snapshot of a distributed system} consists of an event per process. A snapshot of the system is a \textit{consistent snapshot} if every pair of events in it are concurrent with each other. Predicate detection involves identifying a consistent snapshot of the system where a predicate or condition $\pred$ becomes true. $\pred$ can correspond to a violation of a system requirement.
A consistent snapshot of the system where $\pred$ is true is defined as a \textbf{valid snapshot} of the system.

Predicate $\pred$ is a condition defined over the variables of more than one process, so we also refer to it as a global predicate. A local predicate is a condition defined over the variables of a single process. Specifically, local predicate $\pred_i$ of a process $i$ is a condition defined over the variables of process $i$.
The global predicate $\pred$ is called a conjunctive predicate if it is of the form $\bigwedge\pred_i$, where $\pred_i$ is a local predicate at process $i$.



\subsection{Hybrid Logical Clocks\cite{HLC}}
\label{sec:hlc}
In this section, we present Hybrid Logical Clocks (HLC). An HLC timestamp of an event $e$ ($hlc.e$) is an ordered pair $\br{l.e, c.e}$.
Intuitively, if event $e$ happened at process $i$, then $l.e$ captures the maximum physical clock value (of any process) that process $i$ was aware of when event $e$ happened. When two events with the same $l$ value have a happened before relation between them, $c.e$ is a counter that is used to capture causality. 

Let $i$ be the process where event $e$ occurred and let $pt.i$ denote the physical clock value of $i$ when $e$ occurred. Let $\br{l.i,c.i}$ be the timestamp of the last event (before $e$) on process $i$. 
If $e$ is a send event or a local event then, $l.e$ is set to max of $l.i$ and $pt.i$. If the value of $l.e$ equals $l.i$ then $c.e$ is set to $c.i+1$. Otherwise, it is reset to $0$. 
If $e$ is a receive event where it receives a message $m$ with timestamp $\br{l.m,c.m}$, then $l.e$ is set to max of $l.i$, $pt.i$ and $l.m$. If $l.e$ equals $l.m$ or $l.i$, then the $c$ value is used to capture the causal relation between $e$ (message receive event in this case), the event corresponding to sending of $m$ and the previous event on process $i$.
The full algorithm for HLC is shown in Algorithm \ref{alg:HLC}.
\begin{algorithm}
\algnewcommand\algorithmforsendandlocal{\textbf{Send/Local Event}}
\algnewcommand\SENDORLOCALEVENT{\item[\algorithmforsendandlocal]}

\algnewcommand\algorithmforreceive{\textbf{Receive Event of message \em{m}}}
\algnewcommand\RECEIVEEVENT{\item[\algorithmforreceive]}
\caption{HLC Algorithm from \cite{HLC}}
\label{alg:HLC}
\begin{algorithmic}[1] 
\small
\SENDORLOCALEVENT
\State {$l$'$.i := l.i$} 
\State {$l.i := max(l$'$.i, pt.i)$ //tracking maximum time event, $pt.i$ is physical time at $i$}
\State {If ($l.i = l$'$.i$) then $c.i := c.i + 1$  //tracking causality}
\State{Else $c.i := 0$}
\State {Timestamp event (and {message for send event)} with $l.i$,$c.i$}

\RECEIVEEVENT
\State {$l$'$.i := l.i$}
\State {$l.i := max(l$'$.i, l.m, pt.i)$ // $l.m$ is $l$ value in the timestamp of the message received}
\State {If ($l.i = l$'$.i = l.m$) then $c.i := max(c.i,c.m) + 1$}
\State {Elseif ($l.i = l$'$.i$) then $c.i := c.i + 1$}
\State {Elseif ($l.i = l.m$) then $c.i := c.m + 1$}
\State {Else $c.i := 0$}
\State {Timestamp event with $l.i$,$c.i$}
\end{algorithmic}
\end{algorithm}

A key property of HLC is that if event $e$ happened before event $f$, then $hlc.e < hlc.f$, where the $<$ relation is defined in a lexicographic manner. Specifically. 

\begin{tabbing}

\hspace*{5mm} \= $(e \rightarrow f)\ \ \Rightarrow \ \  (hlc.e < hlc.f)$, \\
where\\
\> 
$hlc.e < hlc.f$
iff
$l.e < l.f \vee ((l.e = l.f) \wedge (c.e < c.f))$
\end{tabbing}


An implication of this is that if $hlc.e = hlc.f$, then event $e$ is concurrent with event $f$, i.e., $e || f$.

Another key property of HLC is that the partial synchronization bound $\epsilon$ applies not only to difference in physical clocks but also to the difference in the $l$ values. That is, if the physical clocks of processes are bounded to be within $\epsilon$ of each other, then the $l$ values of different processes also differ by at most $\epsilon$. Finally, the $c$ value is bounded by a maximum value $c_{max}$. (The specific value of $c_{max}$ is not relevant to the paper. We only use the fact that it exists.)


\section{Predicate Detection with HLC: Trade-off in False Positives and Negatives}
\label{sec:predDetHLCwithgamma}
\subsection{Basic Approach} 
\label{sec:HLCBasedPredOriginal}

In this section, we describe the HLC based monitoring from \cite{PredDetHLC_ICDCN}, where the monitor receives information about all events that affect satisfaction of $\pred$ being monitored.  
For simplicity, assume that the monitor processes all the events in an increasing HLC timestamp  order. 
When the monitor processes the event at timestamp $t$, it evaluates $\pred$ in a global snapshot at $t$, where the HLC timestamp of each process is $t$.
By the property of HLC, this snapshot at $t$ is guaranteed to be consistent.
%
%

\begin{figure*}
\begin{center}
\begin{adjustbox}{width=\textwidth,trim=10 0 0 0}
{\begin{tikzpicture} []
\draw [very thin] (0.5,1.1)--(3,1.1);
\node at (0.5,1.1) [below,below] {\scriptsize{$p_0$}};
\node at (0.5,1.4) [below,right]{ \fontsize{4}{5} $\br{1,0}$};
\node at (0.8,0.9) [below,right]{\scriptsize{$e_1$}};
\draw (0.9,1)--(0.9,1.3);%
\draw [very thin] (0.5,0.5)--(3,0.5);
\node at (0.5,0.5) [below,below] {\scriptsize{$p_1$}};
\node at (2,0.9) [below,right]{ \fontsize{4}{5} $\br{7,0}$};
\node at (2.3,0.3) [below,right]{\scriptsize{$f_1$}};
\draw (2.3,0.3)--(2.3,0.7);
\node at (0,0) [right,right]{\tiny{\begin{tabular}{l}(a) Monitor considers the state at \br{1,0}  \end{tabular}}
};
%
%
%
\draw [very thin] (4,1.1)--(7,1.1);
\node at (4,1.1) [above,left] {\scriptsize{$p_0$}};
\node at (3.6,1.4) [below,right]{ \fontsize{4}{5} $\br{1,0}$};
\node at (4.1,0.9) [below,right]{\scriptsize{$e_1$}};
\draw (4.3,1)--(4.3,1.3);%
\node at (4.5,1.4) [below,right]{ \fontsize{4}{5} $\br{3,0}$};
\node at (4.8,0.9) [below,right]{\scriptsize{$e_2$}};
\draw (4.8,1)--(4.8,1.3);
%
\draw [very thin] (4,0.5)--(7,0.5);
\node at (4,0.5) [above,left] {\scriptsize{$p_1$}};
\node at (5.6,0.9) [below,right]{ \fontsize{4}{5} $\br{7,0}$};
\node at (6,0.3) [below,right]{\scriptsize{$f_1$}};
\draw (6,0.3)--(6,0.7);
\node at (5.3,0) {\tiny{\begin{tabular}{l}(b) Monitor considers the state at \br{3,0} only
\end{tabular}}};
%
%
%
\draw [very thin] (7.8,1.1)--(12,1.1);
\node at (7.8,1.1) [above,left] {\scriptsize{$p_0$}};
\node at (7.5,1.4) [below,right]{ \fontsize{4}{5} $\br{1,0}$};
\node at (8,0.9) [below,right]{\scriptsize{$e_1$}};
\draw (8.1,1)--(8.1,1.3);%
\draw [dashed] (8.1,1.2)--(9.4,1.2);
\draw (9.4,1)--(9.4,1.2);%
\node at (9.05,0.9) [below,right]{\scriptsize{$e_1'$}};
\node at (9.2,0.9) [below,right]{ \fontsize{4}{5} $\br{6,c_{max}}$};
\node at (8.1,1.5) [below,right]{ \fontsize{4}{5} $\br{3,0}$};
\node at (8.5,0.9) [below,right]{\scriptsize{$e_2$}};
\draw (8.6,1)--(8.6,1.4);
\draw [dashed] (8.6,1.3)--(9.7,1.3);
\draw [dashed] (9.7,1.0)--(9.7,1.3);%
\node at (9.5,1.5) [below,right]{\scriptsize{$e_2'$}};
\node at (9.65,1.5) [below,right]{ \fontsize{4}{5} $\br{8,c_{max}}$};
%
\draw [very thin] (7.8,0.5)--(12,0.5);
\node at (7.8,0.5) [above,left] {\scriptsize{$p_1$}};
\node at (8.7,0.3) [below,right]{ \fontsize{4}{5} $\br{7,0}$};
\node at (9.6,0.3) [below,right]{\scriptsize{$f_1$}};
\draw (9.6,0.3)--(9.6,0.7);
\draw [dashed] (9.6,0.6)--(11,0.6);
\draw [dashed] (11,0.3)--(11,0.6);%
\node at (11,0.3) [below,right]{\scriptsize{$f_1'$}};
\node at (10.75,0.7) [below,right]{ \fontsize{4}{5} $\br{12,c_{max}}$};
\node at (9.9,0) {\tiny{\begin{tabular}{l}(c) With $\epsilon$-extension: Monitor considers the state at \br{1,0} and \br{3,0}\end{tabular}}};
\end{tikzpicture}}
\end{adjustbox}
\end{center}
\caption{\small{Analyzing the state of $p_0$ considered by the monitor  when evaluating the global predicate at \br{7,0}. ($\epsilon=5$ in the system) (a) \& (b) consider a monitor that uses HLC, (c) considers a monitor that uses HLC with $\epsilon$-extension.
}}
\label{fig:concurrencyHLCEpsExt}
\end{figure*}
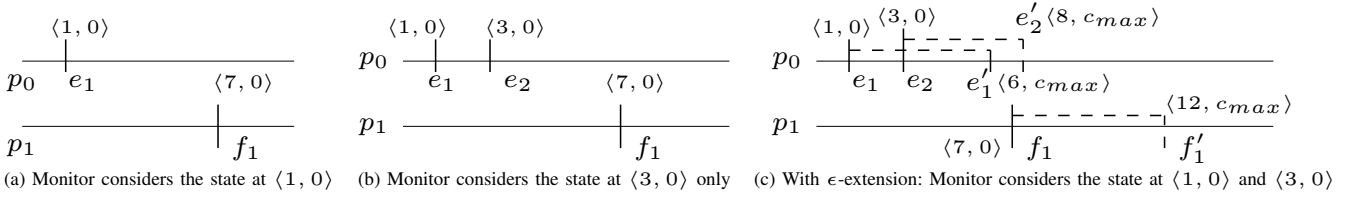

The key property that this monitor uses is that the state of each process does not change between events.
For example, consider the execution in Figure \ref{fig:concurrencyHLCEpsExt}(a) and the timestamp \br{7,0}. The monitor needs to evaluate the predicate at both the processes at \br{7,0}. For process $p_1$, this is trivial.
{For process $p_0$, the monitor evaluates the state at \br{7,0} by evaluating $p_0$'s state at \br{1,0}; this works because the state of the process does not change without an event and there is no event for $p_0$ between \br{1,0} and \br{7,0}.}
Essentially, what this HLC monitor does is compute all consistent snapshots corresponding to every event $e$ using the closest preceding event $f$ (i.e., event with largest timestamp less than or equal to $hlc.e$) on every other process.
{Stated another way, this HLC monitor does two things when processing an event $e_1$ on process $i$: (1) it \emph{ends} the current state of process $i$ that was in effect before $hlc.e_1$ and (2) \emph{introduces} a new state for process $i$ that is in effect from $hlc.e_1$ until the $hlc.e_2$ where $e_2$ is the next event on process $i$.}
For more details, as well as handling various complexities such as the monitor receiving events out of order, please refer to  \cite{PredDetHLC_ICDCN}. 

\subsection{False Negatives with HLC}
\label{sec:falsenegativesWithHLC}


A disadvantage with the predicate detection approach discussed above is that it suffers from false negatives.
In particular, lets reexamine what happens when it computes the global snapshot for a given timestamp $t$ for an event $e$ on process $i$.
When it considers a process $j \neq i$, it will only use the state of process $j$ from the most recent preceding event $f$ {(i.e., the event with largest $HLC$ timestamp that is less than or equal to $t=hlc.e$)}.
This is not a problem if event $f$ occurs more than $\epsilon$ before $t$, but it is problematic if $f$ occurs within $\epsilon$ of $t$.

For example, consider Figure \ref{fig:concurrencyHLCEpsExt}(b) where $\epsilon =5$ with a new event $e_2$ on $p_0$ at timestamp \br{3,0} which is  within $\epsilon$ of \br{7,0}.
When the HLC monitor computes the global snapshot at timestamp \br{7,0}, for process $p_0$, it will only consider the state of process $p_0$ after event $e_2$ even though the period prior to event $e_2$ also lies within $\epsilon$ of \br{7,0} and thus is concurrent with event $f_1$.
To ensure there are no false negatives, the monitor needs to consider all possible states of $p_0$ that can be concurrent with $t$; namely any state of $p_0$ within $\epsilon$ of $t$.

\subsection{Eliminating False Negatives with $\epsilon$ extension}
\label{sec:eliminateFNwithEpsExt}
As discussed in Section \ref{sec:falsenegativesWithHLC}, HLC based monitor misses consistent snapshots consisting of events $e_1$ and $f_1$ where there exists an event $e_2$ such that $hlc.e_1 < hlc.e_2 < hlc.f_1$ and {$l.f_1 - l.e_2 < \epsilon$}. To overcome this limitation, we introduce the notion of $\epsilon$-extension, where the effect of event $e_1$ (in previous sentence) is extended by $\epsilon$. 
Specifically, for an event $e$ if $hlc.e$ is \br{l.e, c.e}, we define $hlc.e+\epsilon$ to be $\br{l.e+\epsilon, c_{max}}$, where $c_{max}$ is the maximum possible value of $c$. Extending by $\epsilon$ leads to multiple possible values for a process where the extended intervals overlap.


To illustrate this, let $x_i$ be a variable of process $i$ used in $\pred$ that is being monitored,
and suppose $x_i$ is set to $v_0$ at timestamp $t_0$ and then changed to $v_1$ at timestamp $t_1$.
As we observed earlier, the HLC monitor would use the event at timestamp $t_1$ to perform two actions: (1) remove $v_0$ as the value of $x_i$  and (2) add $v_1$ as the value of $x_i$.
With $\epsilon$-extension, at timestamp $t_1$, we only perform the second action of adding $v_1$ as a possible value for $x_i$. At timestamp $t_1+\epsilon$, we perform the first action of removing $v_0$ as a possible value of $x_i$. 
This extends the effective interval where $x_i$ has value $v_0$ from $\closeopen{t_0,t_1}$ to $\closeopen{t_0, t_1+\epsilon}$.
Note that we explicitly must consider multiple possible values, in this case $v_0$ and $v_1$, for $x_i$ for any snapshot in the time interval $\closeopen{t_1, t_1+\epsilon}$.
Stated more generally, with $\epsilon$-extension, we turn each event $e$ into two events, $e$ and $e'$, where event $e$ has its original timestamp $t$ and event $e'$ has timestamp $t+ \epsilon$.
At event $e$, we perform the second action where we add a new possible value for a variable.
At event $e'$, we perform the first action where we remove a possible value for a variable.


As an illustration, consider the execution in Figure \ref{fig:concurrencyHLCEpsExt}, where we have events $e_1$, $f_1$ and $e_2$ as shown. Effect of event $e_1$ would be in the interval \closeopen{\br{1,0},\br{3,0}} (c.f. Figure \ref{fig:concurrencyHLCEpsExt}b). The $\epsilon$-extension (c.f. Figure \ref{fig:concurrencyHLCEpsExt}c) will change it to interval \closeopen{\br{1,0},\br{8,c_{max}}}. The effect of event $e_2$ is in the interval $\closeopen{\br{3,0},\br{\infty,-}}$. 
Thus, in the interval between \br{3,0} and \br{8,c_{max}}, we consider both states corresponding to events $e_1$ and $e_2$. 
\color{black}

While having to consider multiple values will increase complexity and decrease efficiency, we expect that the number of values are likely to be small given that we are only extending effective intervals by $\epsilon$. 
For example, in the above scenario, multiple values of $x_i$ are only considered in the interval $\closeopen{t_1, t_1 +\epsilon}$.
Also, it is expected that such multiple values would need to be considered only for a small subset of processes. 
For example, not all processes would have multiple permitted values in the interval $\closeopen{t_1, t_1+\epsilon}$. 

Even better, for many predicates $\pred$, it is possible to eliminate multiple values altogether. 
For example, consider the conjunctive predicate $\pred = \bigwedge \pred_i$.
Clearly, for any interval where $\pred_i$ can be both true and false, we can just record that $\pred_i$ is true.
For example, if $\pred_i$ is set to true at $t_0$ and then false at $t_1$, then in the interval $\closeopen{t_0,t_1+\epsilon}$, we would only record that $\pred_i$ is true and only begin recording that $\pred_i$ is false at timestamp $t_1+\epsilon$.
Beyond conjunctive predicates, we can also handle predicates of the form $\Sigma x_i > C$ and $\Sigma x_i < C$.
In the first case, we choose the maximum value of $x_i$; in the second case, we choose the minimum value of $x_i$.
For predicates of the form $x_1 < x_2$, we can choose the largest value of $x_1$ and smallest value of $x_2$.
The first case $\Sigma x_i > C$ includes violation of mutual exclusion, where $x_i=1$ means process $i$ is accessing the shared resource and we set $C=1$.

{Observe that the $\epsilon$-extension does not account for messages. Therefore, it will lead to false positives. Furthermore, instead of detecting the given predicate $\pred$, if we detect a slightly weaker predicate, it will increase the rate of false positives. Allowing false positives will enable us to detect more complex predicates. For example, if the predicate we want to detect is $C_1 < \Sigma x_i < C_2$, we can split this predicate into two separate predicates, one for the upper bound and one for the lower bound, both of which we can detect efficiently.
We may increase the false positive rate with respect to the original predicate if there are no choices of $x_i$ that simultaneously satisfy both bounds. We discuss in Section \ref{sec:FPVFNVimplications} how the false positive rate can be effectively managed. }

\color{black}

\subsection{Reducing False Positives with $\gamma$-extension}
\label{sec:reduceFPVwithGamma}

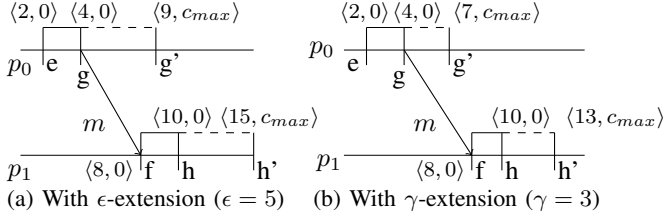
\begin{figure}
\begin{center}
\begin{tikzpicture}
\draw [very thin] (0,2.0)--(3.1,2.0);
\draw [very thin] (0,0.6)--(3.1,0.6);
\node at (0,2) [above,below] {$p_0$};
\node at (0,0.6) [above,below] {$p_1$};
\node at (-0.4,2.5) [below,right]{ \fontsize{8}{10} $\br{2,0}$};
\node at (0.1,1.8) [below,right]{ \fontsize{8}{10} e};
\draw (0.3,1.8)--(0.3,2.3);
\node at (0.4,2.5) [below,right]{ \fontsize{8}{10} $\br{4,0}$};
\node at (0.5,1.6) [below,right]{ \fontsize{8}{10} g};
\draw (0.8,1.8)--(0.8,2.3);
\draw [very thin] (0.3,2.3)--(0.8,2.3);
\node at (1.5,2.5) [below,right]{ \fontsize{8}{10} $\br{9,c_{max}}$};
\node at (1.6,1.8) [below,right]{ \fontsize{8}{10} g'};
\draw (1.8,1.8)--(1.8,2.3);
\draw [dashed] (0.8,2.3)--(1.8,2.3);
%
%
\node at (0.6,0.4) [below,right]{ \fontsize{8}{10} $\br{8,0}$};
\node at (1.4,0.4) [below,right]{ \fontsize{8}{10} f};
\draw (1.6,0.4)--(1.6,0.9);
\node at (1.5,1.1) [below,right]{ \fontsize{8}{10} $\br{10,0}$};
\node at (1.9,0.4) [below,right]{ \fontsize{8}{10} h};
\draw (2.1,0.4)--(2.1,0.9);
\draw [very thin] (1.6,0.9)--(2.1,0.9);
\node at (2.4,1.1) [below,right]{ \fontsize{8}{10} $\br{15,c_{max}}$};
\node at (2.9,0.4) [below,right]{ \fontsize{8}{10} h'};
\draw (3.1,0.4)--(3.1,0.9);
\draw [dashed] (2.1,0.9)--(3.1,0.9);
\draw [->] (0.8,2)--(1.6,0.6);
\node at (0.7,1)[right,right] {$m$};
\node at (-0.3,0)[right,right] {\small(a) With $\epsilon$-extension ($\epsilon=5$)};
%
%
%
%
\node at (4.3,2) [above,left] {$p_0$};
\node at (4.4,0.5) [above,left] {$p_1$};
\draw [very thin] (4.3,2.0)--(7.5,2.0);
\draw [very thin] (4.3,0.6)--(7.5,0.6);
%
\node at (4,2.5) [below,right]{ \fontsize{8}{10} $\br{2,0}$};
\node at (4.1,1.8) [below,right]{ \fontsize{8}{10} e};
\draw (4.6,1.8)--(4.6,2.3);
\node at (4.7,2.5) [below,right]{ \fontsize{8}{10} $\br{4,0}$};
\node at (4.7,1.6) [below,right]{ \fontsize{8}{10} g};
\draw (5.1,1.8)--(5.1,2.3);
\draw [very thin] (4.6,2.3)--(5.1,2.3);
\node at (5.5,2.5) [below,right]{ \fontsize{8}{10} $\br{7,c_{max}}$};
\node at (5.5,1.8) [below,right]{ \fontsize{8}{10} g'};
\draw (5.7,1.8)--(5.7,2.3);
\draw [dashed] (5.1,2.3)--(5.7,2.3);
%
%
\node at (5,0.4) [below,right]{ \fontsize{8}{10} $\br{8,0}$};
\node at (5.8,0.4) [below,right]{ \fontsize{8}{10} f};
\draw (6,0.4)--(6,0.9);
\node at (6,1.1) [below,right]{ \fontsize{8}{10} $\br{10,0}$};
\node at (6.2,0.4) [below,right]{ \fontsize{8}{10} h};
\draw (6.4,0.4)--(6.4,0.9);
\draw [very thin] (6,0.9)--(6.4,0.9);
\node at (7,1.1) [below,right]{ \fontsize{8}{10} $\br{13,c_{max}}$};
\node at (6.9,0.4) [below,right]{ \fontsize{8}{10} h'};
\draw (7.1,0.4)--(7.1,0.9);
\draw [dashed] (6.4,0.9)--(7.1,0.9);
\draw [->] (5.1,2)--(6,0.6);
\node at (5.1,1)[right,right] {$m$};
\node at (5.8,0) {\small(b) With $\gamma$-extension ($\gamma=3$)};
\end{tikzpicture}
\end{center}
\caption{\small{Reducing false positives with $\gamma$-extension}}
\label{fig:fpvfnvGammaExt}
\end{figure}

The $\epsilon$-extension eliminates false negatives at the cost of introducing false positives. In particular, the $\epsilon$-extension  allows the state of a process to be extended by $\epsilon$ into the future even if causality shows that such an extension is impossible. 
As an illustration, in Figure \ref{fig:fpvfnvGammaExt}(a) , event $e$ happened before $g$ which happened before $f$. However, the approach of $\epsilon$-extension would incorrectly allow the state of $p_0$ at \br{2,0} to be considered by the monitor at timestamp \br{8,0}.
In other words, the monitor considers $e$ and $f$ to be concurrent even though they are not, resulting in a false positive. 

If the intervals in Figure \ref{fig:fpvfnvGammaExt}(a) were extended for a shorter duration, say $\gamma < \epsilon$, then the rate of false positives will decrease at the cost of reintroducing false negatives. For example, for $\gamma=3$, the state of $p_0$ at \br{2,0} will never be considered at timestamp \br{8,0}
(c.f. Figure \ref{fig:fpvfnvGammaExt}(b)). Thus, it will not be part of the false positives. However, for $\gamma=3$, in the absence of communication (even when $e\not\rightarrow f$), the snapshot consisting of $e$ and $f$ will be discarded, thereby resulting in a false negative.

The parameter $\gamma$ provides a mechanism to control the rate of false positives and false negatives. 
In the next section, we analyze the false positives/negatives for different values of $\gamma$ and other system parameters. 
%
%
Specifically,  we evaluate the effectiveness of performing predicate detection using  $\gamma$-extension through experimental analysis; that is, we compute the precision (1 $\!-\!$ false positive rate) and recall (1 $\!-\!$ false negative rate)  of a monitor that uses $\gamma$-extension to perform conjunctive predicate detection.

\subsection{Analyzing False Positives and Negatives with $\gamma$ extension}
\label{sec:gammaFPVAnalysis}

\subsubsection{Experimental Setup}\label{sec:gammaExtExpSetup}
To analyze the effectiveness of $\gamma$-extension in performing predicate detection, we simulate a distributed system of 10 processes. 
Although this analysis is for the case of conjunctive predicates, we note that the approach is general enough to be applied for other predicates. We use conjunctive predicates because evaluating the ground truth (identifying all valid snapshots) is feasible using \cite{garg_conj_pred}. And, this ground truth is essential to compute false positives and negatives.

In this work, each process has a physical clock $pt$ and a hybrid logical clock associated with it. Each process $i$ is also associated with a boolean variable $v_i$. The processes execute in a round robin fashion and each process executes a million times. Each time a process executes it advances its physical clock with a certain probability such that the physical clock value of any two processes differ by at most $\epsilon$.  When a process advances its clock, it sends a message with a probability $\alpha$ to a uniformly randomly chosen process. Based on the value of $\delta$-message delay and time at which a message was sent, the process receives any message that was sent to it if it is ready to be delivered. If the value of the variable $v_i$ is false the process sets its value to true with a probability $\beta$. 
If $v_i$ is true, it remains true for a duration of length $\ell$ (counted in terms of physical clock of $i$). Then, it is set to false.  
The process updates its hybrid logical clock value every time it sends or receives a message, and whenever it changes the value of $v_i$ based on the HLC algorithm presented in Section \ref{sec:hlc}.

Each process $i$ reports every duration for which $v_i$ was true as an interval to a common monitoring process. Each interval consists of the HLC timestamp when $v_i$ became true and the HLC timestamp when $v_i$ became false after that.
{The monitor applies $\gamma$-extension by adding $\gamma$ to the interval-end timestamp since it is always advantageous to choose $v_i$ to be true whenever there is a choice.
The monitor then reports all snapshots where $\bigwedge v_i$ becomes true (i.e., when intervals of all processes overlap).}

We extend the predicate detection algorithm in \cite{garg_conj_pred} to identify the ground truth with two modifications.
First, we replace Vector Clocks with Hybrid Vector Clocks\cite{HVC} to account for clock synchronization.
Second, we modified the algorithm so that it continues the detection process until all valid snapshots are identified.
Specifically, after finding a valid snapshot, we look for the next non-overlapping valid snapshot.

We compute the number of true positives, false positives, and false negatives by comparing the results returned by both monitoring solutions.
False positives are snapshots $S$ identified using $\gamma$-extension where neither $S$ nor any overlapping snapshot exists in the ground truth.
False negatives are snapshots $S$ identified using the ground truth  
where neither $S$ nor any overlapping snapshot is found using $\gamma$-extension.

\subsubsection{Observation}
\label{sec:gammaFPVAnalysisObservation}
To analyze the effectiveness of $\gamma$-extension, 
we computed the true positives, false positives and false negatives reported by the monitor under different settings as we varied the system parameters: $\alpha, \beta, \epsilon, \delta$, and $\ell$ as well as parameter $\gamma$ used by the monitor.
Our precision\footnote{\textbf{Precision} of the monitor is the ratio of the number of \textit{valid} snapshots (i.e., those that are consistent and where $\pred$ is true)  detected by the monitor to the total number of snapshots reported by it.} and recall\footnote{\textbf{Recall} of the monitor is the ratio of the number of valid snapshots detected by the monitor to the number of actual valid snapshots in the system. } results are displayed in Tables \ref{tab:fpvBeta}\footnote{We report the precision as NA when
the monitor does not detect any snapshots and therefore detects no valid snapshots and the precision would be $0/0$. This only happens for small $\gamma$ (for example $\beta=0.02, \gamma=0.1* \epsilon$).}
and \ref{tab:fnvBeta}, respectively .
We observe that irrespective of the underlying system setting, the precision of the monitor decreases and the recall increases as the value of $\gamma$ increases. 
For example, for $\beta=0.045$, as the value of $\gamma$ increases from $0.1 * \epsilon$ to $\epsilon$, the precision of the monitor drops from $0.929$ to $0.295$.
On the other hand, for $\beta=0.045$, as the value of $\gamma$ increases from $0.1 * \epsilon$ to $\epsilon$ the recall of the monitor increases from $0.009$ to $1$.

%
For reasons of space, we relegate the detailed analysis of false positives and negatives to the Appendix. 

Note that in scenarios where the value of clock skew is not known to the processes (or the monitor), the monitor can use a value of $\epsilon$, say $\epsilon_1$, that seems reasonable. If the actual value is $\epsilon_2, \epsilon_2 > \epsilon_1$ then the effectiveness of the monitor can be calculated by our experiments. In fact, this corresponds to the case where $\gamma = \epsilon_1$ and $\epsilon = \epsilon_2$.



\begin{table*}[ht]
\begin{center}
\subfloat[Percentage of Valid Snapshots out of all snapshots detected during Conjunctive Predicate Detection using $\gamma$-extension. Each entry corresponds to precision ({No. of Valid Snapshots detected/Total Snapshots Detected})]{\label{tab:fpvBeta}
\begin{tabular}{|l|l|l|l|l|l|r|}
\toprule
\textbf{Precision} & \multicolumn{1}{l|}{\textbf{$\beta$ = 0.02}} & \multicolumn{1}{l|}{\textbf{$\beta$ = 0.025}} & \multicolumn{1}{l|}{\textbf{$\beta$ = 0.03}} & \multicolumn{1}{l|}{\textbf{$\beta$ = 0.035}} & \multicolumn{1}{l|}{\textbf{$\beta$ = 0.04}} & \multicolumn{1}{l|}{\textbf{$\beta$ = 0.045}} \\
\midrule
\textbf{$\gamma$ = 0.10 * $ \epsilon$} & NA & NA & 0.333 (1/3) & 1.000 (4/4) & 0.739 (17/23) & 0.929 (26/28)\\
\textbf{$\gamma$ = 0.15 * $ \epsilon$} & 0.500 (1/2) & 0.000 (0/3) & 0.529 (9/17)& 0.650 (13/20)& 0.797 (102/128)& 0.743 (113/152)\\
\textbf{$\gamma$ = 0.20 * $ \epsilon$} & 0.500 (2/4)& 0.286 (2/7)& 0.511 (46/90)& 0.582 (46/79)& 0.732 (341/466)& 0.742 (339/457)\\
\textbf{$\gamma$ = 0.25 * $ \epsilon$} & 0.200 (3/15)& 0.190 (4/21)& 0.477 (126/264)& 0.466 (110/236)& 0.688 (778/1131)& 0.706 (771/1092)\\
\textbf{$\gamma$ = 0.50 * $ \epsilon$} & 0.077 (44/572)& 0.070 (42/604)& 0.192 (699/3635)& 0.192 (722/3756)& 0.383 (2756/7188)& 0.384 (2780/7246)\\
\textbf{$\gamma$ = 0.75  * $ \epsilon$} & 0.025 (71/2827)& 0.024 (70/2931)& 0.107 (809/7575)& 0.109 (844/7721)& 0.304 (2877/9473)& 0.305 (2905/9518)\\
\textbf{$\gamma$ = $\epsilon$} & 0.013 (71/5651)& 0.013 (71/5665)& 0.088 (813/9208)& 0.092 (850/9252)& 0.292 (2879/9855)& 0.295 (2907/9859)\\
\hline
\end{tabular}}
\end{center}
%
%
%
%
%
%
%
%
%
\begin{center}
\subfloat[Percentage of Valid Snapshots detected during Conjunctive Predicate Detection using $\gamma$-extension out of all Valid Snapshots in the system. Each entry corresponds to recall ({No. of Valid Snapshots detected/No. of Valid Snapshots in the system})]{\label{tab:fnvBeta}\begin{tabular}{|l|l|l|l|l|l|r|}
\toprule
\textbf{Recall} & \multicolumn{1}{l|}{\textbf{$\beta$ = 0.02}} & \multicolumn{1}{l|}{\textbf{$\beta$ = 0.025}} & \multicolumn{1}{l|}{\textbf{$\beta$ = 0.03}} & \multicolumn{1}{l|}{\textbf{$\beta$ = 0.035}} & \multicolumn{1}{l|}{\textbf{$\beta$ = 0.04}} & \multicolumn{1}{l|}{\textbf{$\beta$ = 0.045}} \\
\midrule
\textbf{$\gamma$ = 0.10 * $ \epsilon$} & 0.000 (0/71)& 0.000 (0/71)& 0.001 (1/813)& 0.005 (4/850)& 0.006 (17/2879)& 0.009 (26/2907)\\
\textbf{$\gamma$ = 0.15 * $ \epsilon$} & 0.014 (1/71)& 0.000 (0/71)& 0.011 (9/813)& 0.015 (13/850)& 0.035 (102/2879)& 0.039 (113/2907)\\
\textbf{$\gamma$ = 0.20 * $ \epsilon$} & 0.028 (2/71)& 0.028 (2/71)& 0.057 (46/813)& 0.054 (46/850)& 0.118 (341/2879)& 0.117 (339/2907)\\
\textbf{$\gamma$ = 0.25 * $ \epsilon$} & 0.042 (3/71)& 0.056 (4/71)& 0.155 (126/813)& 0.129 (110/850)& 0.270 (778/2879)& 0.265 (771/2907)\\
\textbf{$\gamma$ = 0.50 * $ \epsilon$} & 0.620 (44/71)& 0.592 (42/71)& 0.860 (699/813)& 0.849 (722/850)& 0.957 (2756/2879)& 0.956 (2780/2907)\\
\textbf{$\gamma$ = 0.75  * $ \epsilon$} & 1.000 (71/71)& 0.986 (70/71)& 0.995 (809/813)& 0.993 (844/850)& 0.999 (2877/2879)& 0.999 (2905/2907)\\
\textbf{$\gamma$ = $\epsilon$} & 1.000 (71/71)& 1.000 (71/71)& 1.000 (813/813)& 1.000 (850/850)& 1.000 (2879/2879)& 1.000 (2907/2907)\\
\hline
\end{tabular}}
\end{center}
\caption{Precision and Recall when varying $\beta$ - rate at which the local predicate becomes true at a process (Default values: n = 10, $\epsilon$ = 100, $\alpha$ = 0.1, $\delta$ = 10, $l$ = 1)}
\end{table*}%

\subsection{Implications of False Negatives/Positives}
\label{sec:FPVFNVimplications}
From Tables \ref{tab:fpvBeta} and \ref{tab:fnvBeta}, we see that the use of $\gamma$-extension can suffer from high false positives and/or false negatives. 
We now discuss how we can cope with false positives and false negatives. 

We first consider the case where we only have false positives.
We expect that the violations are likely to be rare. This is due to the fact that we generally deploy programs that are \textit{mostly correct}.  As an illustration, suppose that the likelihood of an error in a given time unit (say 1 $second$) is $0.1\%$ and the false positive rate is 90\%.
That means that the likelihood that the monitor reports an error in the given time unit is $1\%$. In this setting,
{we can use a second monitor that is accurate but expensive to analyze these computation slices to determine which errors are real.}
The key observation is that, in this case, 99\% of the computation would not be analyzed with the expensive monitor. We use this idea to develop a two-layered monitor in Section \ref{sec:twolayeredMonitor}. 


%

We next consider the case of false negatives.
While false negatives are not ideal, in many situations, they may be acceptable.
For example, if most false negatives are latent errors  (rarely manifest in practice but can be detected with an accurate monitor), even a 90\% false negative rate (i.e., only 1 in 10 latent errors is identified) means the error, on average, would take roughly 10 times as long to detect.
Furthermore, we typically only resort to using a monitor that has false negatives when the resource requirements of more accurate monitors, such as the one from \cite{garg_conj_pred} which requires $O(n)$ size clocks, make deploying the more accurate monitor practically impossible. 
In such circumstances, we must use the most accurate monitor that can be deployed.


\color{black}

\section{Predicate detection using SMT solvers}
\label{sec:predDetUsingSMT}

In this section, we summarize the approach in \cite{MonitoringUsingSMT}. {A modification of this monitor becomes the second layer in our proposed monitor.}
In this approach, similar to the HLC based monitor, the processes report changes in the variables involved in the predicate \pred being monitored. Based on this, a set of constraints are created. For example, if process $i$ reports that the value of $x_i$ was set to $val$ at HLC timestamp $t_0$ and changed again at HLC timestamp $t_1$, then the monitor creates the constraint

\indent\indent\indent$(\br{l_i, c_i} \geq t_0) \wedge 
 (\br{l_i, c_i} < t_1) \ \ \ \Rightarrow \ \ \ x_i = val $

In \cite{MonitoringUsingSMT}, the processes also report timestamps corresponding to send and receive of messages. For example, if message $m$ is sent at timestamp $t_0$ by process $i$ and received at timestamp $t_1$ by process $j$ then the monitor adds the constraint that 

\indent\indent\indent\indent\indent\indent$(\br{l_j,c_j} \geq t_1) \Rightarrow (\br{l_i,c_i} > t_0)$

Note that the above constraint captures that if the timestamp of process $j$ reflects message being received then the timestamp of process $i$ must reflect that the message has been sent. Additionally, the monitor creates other constraints such as timestamps of different processes must be within $\epsilon$ of each other {(to ensure that the timestamps correspond to concurrent events that form a consistent snapshot)} and that the predicate $\pred$ must be true in the snapshot. 

These constraints are then fed to an SMT solver (Z3 \cite{Z3} was used in \cite{MonitoringUsingSMT}).
The solver will return a model that corresponds to a consistent snapshot in the run where the predicate (violation) is true. If the constraints are not satisfiable, meaning there is no valid snapshot (i.e., no violation) in the run, the solver will return unsat.

While predicate detection using SMT solvers guarantees the absence of false positives and false negatives, the drawback of this monitoring approach is the high computation time required to perform the detection.
If the monitor takes too long, it cannot be used in an online setting as it will not be able to keep up with the application. The two-layered approach proposed in the next section aims to use HLC based monitor from Section \ref{sec:reduceFPVwithGamma} and the SMT based monitor in \cite{MonitoringUsingSMT} to achieve efficient and accurate predicate detection. 


\section{Two-Layered Monitoring Approach}
\label{sec:twolayeredMonitor}

In this section, we present our two-layered monitoring approach that is both accurate and efficient.
The first layer uses HLC based monitor with $\gamma$-extension, and the second layer uses SMT solvers. 
We invoke the SMT based monitor (which is accurate but inefficient) only if the HLC based monitor with $\gamma$ extension (which is inaccurate but efficient) identifies the possibility that the predicate $\pred$ of interest is likely to be true.
Specifically, the combined monitor works as follows:

\begin{itemize}
    \item Similar to the monitoring approach discussed in Section \ref{sec:predDetUsingSMT}, each process reports changes of variables involved in  predicate $\pred$ to the monitor. It also reports to the monitor the timestamps of messages that are sent and received. 
    \item HLC based monitor with $\gamma$-extension uses the information about variable changes to determine if $\pred$ is true. Note that if $\gamma = \epsilon$, this approach suffers from only false positives. For $\gamma < \epsilon$, it may suffer from false positives and negatives. 
    \item The SMT based monitor creates constraints as discussed in Section \ref{sec:predDetUsingSMT}. These constraints are partitioned into \textit{windows} based on the timestamps of the corresponding events {(message send/receive events and events where the value of a variable changes)}. 
    For example, if $w$ is the length of the window, then the windows correspond to timestamps, $[0..w+\epsilon]$, $[w..2w+\epsilon]$, etc. The overlap of $\epsilon$ is added to ensure that we do not miss snapshots that cross window boundaries. In this paper, we use $w=\epsilon$.
    \item If the monitor based on HLC with $\gamma$-extension finds a consistent snapshot where $\pred$ is satisfied,
    then the SMT solver is invoked on corresponding window.  Otherwise, the constraints from that window are discarded. 
    \item We batch the windows on which the SMT solver is invoked. This means the SMT solver is invoked less often but has to deal with multiple windows at once. 
    
\end{itemize}

\color{black}

The recall of the two-layered monitor depends on the value of $\gamma$ used in the filtering layer.
Based on our discussion in Section \ref{sec:gammaFPVAnalysisObservation}, 
if $\gamma=\epsilon$ in the filtering layer, then the two-layered monitor will have perfect recall.
{If the false positive rate of $\epsilon$-extension is too high thereby resulting in an inability to run the second layer of the monitor efficiently, we may have to use  $\gamma< \epsilon$ thereby sacrificing perfect recall.}
In this case, the two-layered monitor may suffer from false negatives.

Irrespective of the value of $\gamma$ used, the two-layered monitor will have perfect precision, i.e. no false positives, because the solver in the second layer will verify and eliminate all false positives.
In general, using HLC with $\gamma$-extension as a filtering layer will reduce the number of times the monitor has to invoke the solver. This is due to the fact that if the filtering layer does not detect a violation in a window, then the monitor does not have to invoke the solver to check that window.


{To analyze the effect of the two-layered approach, we compare it with the single-layered approach in \cite{MonitoringUsingSMT}.}

\subsection{Evaluating Efficiency of the Two Layered Monitor}
\label{sec:hlcgammaSMTExpeAnalysis}

\subsubsection{Application based on time division multiplexing.} To evaluate the effectiveness of the two-layered monitor, we use it to monitor an application (considered in \cite{MonitoringUsingSMT}) that uses time division multiplexing to ensure exclusive access to a shared resource. The application has the same setup discussed in Section \ref{sec:gammaExtExpSetup}, except for the parameters $\beta$ and $\ell$. In this application, the value of the variable $v_i$ at process $i$ denotes if the process $i$ is accessing the shared resource ($v_i=true$) or not ($v_i=false$).

The basic intuition of time division multiplexing is as follows. Process $0$ accesses the resource in interval \closeopen{0,T}, process $1$ accesses it in the interval \closeopen{T, 2T}, and so on. After the last process accesses the resource, process $0$ can access it again. And, the cycle repeats. To account for clock skew and avoid simultaneous access, the access-windows are changed to $\closeopen{0, T-\epsilon}$, $\closeopen{T, 2T-\epsilon}$ and so on. However, we introduce an error so that the time to release the resource is changed with probability of 10\%. This will cause some processes to continue accessing the resource while the next process in the sequence starts accessing the resource. 
Thus, the predicate $\pred$ of interest captures that two or more processes are accessing the resource simultaneously. 

\subsubsection{Two-layered Monitoring Setup} 
In the two layered monitoring setup, we treat every $100$ windows (recall that each window $w$ is of duration $\epsilon$) of the application run  as a \textit{batch}. 
At the end of every batch, the monitor first performs predicate detection using HLC with $\gamma$-extension by processing all the intervals\footnote{Recall that each interval reported by a process $i$ corresponds to a duration for which $v_i$ was true} reported by the processes in the current batch. For every snapshot detected using HLC with $\gamma$-extension, the monitor marks the corresponding window in the batch. Then the monitor invokes the solver to process all the \textit{marked} windows in the batch. Specifically, the monitor generates constraints corresponding to all the marked windows in the batch and feeds it to the solver. It records the total time taken by the solver to evaluate them. 
The monitor then invokes the solver (again) to process all the windows (marked and unmarked) in the batch and records the total time taken by the solver. 

We apply the two-layered monitor in the time division multiplexing protocol based application varying  the parameters $\alpha,\delta,\epsilon$ in the application and $\gamma$ in the monitor
where the monitor is trying to detect if the processes ever violate the exclusive access requirement i.e., if two or more processes access the shared resource simultaneously. 
For each setting, we obtain the overall time taken by the solver to check all the windows in each batch (this corresponds to the time taken by the solver in the single-layered monitor \cite{MonitoringUsingSMT}) and the overall time taken by the solver to check only the windows marked by HLC with $\gamma$-extension in each batch (this corresponds to the time taken by the solver in the two-layered monitor). {We also computed the time taken for predicate detection using HLC with $\gamma$-extension in the filtering layer and observed that it takes around 2 to 3 seconds.} 

\subsubsection{Experimental Results}
We present our experimental results in Figure \ref{fig:SolverTimeToken}. The graphs on the left (Figures \ref{fig:allwindgammadelta}, \ref{fig:allwindgammaeps}, \ref{fig:allwindgammaalpha}) present the time taken (in milliseconds) by the solver in the single-layered monitor \cite{MonitoringUsingSMT}. 
For the same settings, we present the corresponding total time taken (in milliseconds) by the solver in the two-layered monitor 
in the graphs on the right (Figures \ref{fig:markwindgammadelta}, \ref{fig:markwindgammaeps}, \ref{fig:markwindgammaalpha}).

{We note that when $\gamma=\epsilon$, the single layered monitor and the two-layered monitor are both completely accurate with no false positives or false negatives. 
We observe that the two-layered monitor is much more efficient. 
For example, when $n=10,\delta=10,\epsilon=100,\alpha=0.1$ in the application in \ref{sec:hlcgammaSMTExpeAnalysis}, the two layered monitor takes 2 seconds for the first layer and 1.1 seconds for the second layer. By contrast, the single layered monitor takes 81 seconds.
On average, the two-layered monitor reduces the cost by 90\%, with a minimum savings of 85\% across all our parameter settings.
{Furthermore, the maximum time required for the second layer across all our parameter settings is only 2.5 seconds which is comparable to the time required for the first layer. This ensures that the second layer will not become the bottleneck.}

A $\gamma$ extension ($\gamma < \epsilon$) should be used only if the time required for the $\epsilon$-extension monitor is still too high, as it will allow us to reduce the time of monitoring further at the cost of false negatives. 
{
We explore how decreasing $\gamma$ reduces the time required by the second layer.
The three graphs in Figures \ref{fig:markwindgammadelta}, \ref{fig:markwindgammaeps}, \ref{fig:markwindgammaalpha} show a roughly linear decrease in running time for the second layer as a function of $\gamma$ across all three parameters $\alpha$, $\beta$, and $\delta$.
For reasons of space, details of the increase in false positives/negatives are in Tables \ref{tab:fpvAlphaTokenInAppendix} and \ref{tab:fnvAlphaTokenInAppendix} in the Appendix. Note that false positives will be removed by the second layer. 
}
%
%
In general, we observe that the time taken for the two-layered monitoring approach decreases when the message frequency decreases, clock skew decreases, or when the message delay increases.

We note that we considered the problem of conjunctive predicate monitoring in Section \ref{sec:gammaFPVAnalysis} as it forms a basis of all predicates. We considered mutual exclusion in this section as it occurs more frequently in practice. Furthermore, conjunctive predicates cannot be configured easily to have a nonzero but small number of bugs. However, we find similar results in the case of conjunctive predicates when the number of bugs is small. Specifically, we observed the same trend when the two-layered monitor was deployed in the context of conjunctive predicates when the number of valid snapshots in the system was small (<50). For example, for the setting $n=5, \alpha=0.1, \ell=20, \beta=0.004$ where the number of valid snapshots in the system was 38, the time taken by the solver in the single-layered monitor was 16.8s, whereas in the two-layered monitor it took 3.2s. {(We note that having more than 50 bugs in the runs considered here would mean the application is not ready to be deployed; it needs to be tested thoroughly,  not monitored at runtime.)}

\begin{figure*}
\subfloat[Varying $\delta$ and $\gamma$\label{fig:allwindgammadelta} ]{\includegraphics[clip,trim=0.1cm 3.5cm 0.1cm 3.5cm,width=0.5\textwidth]{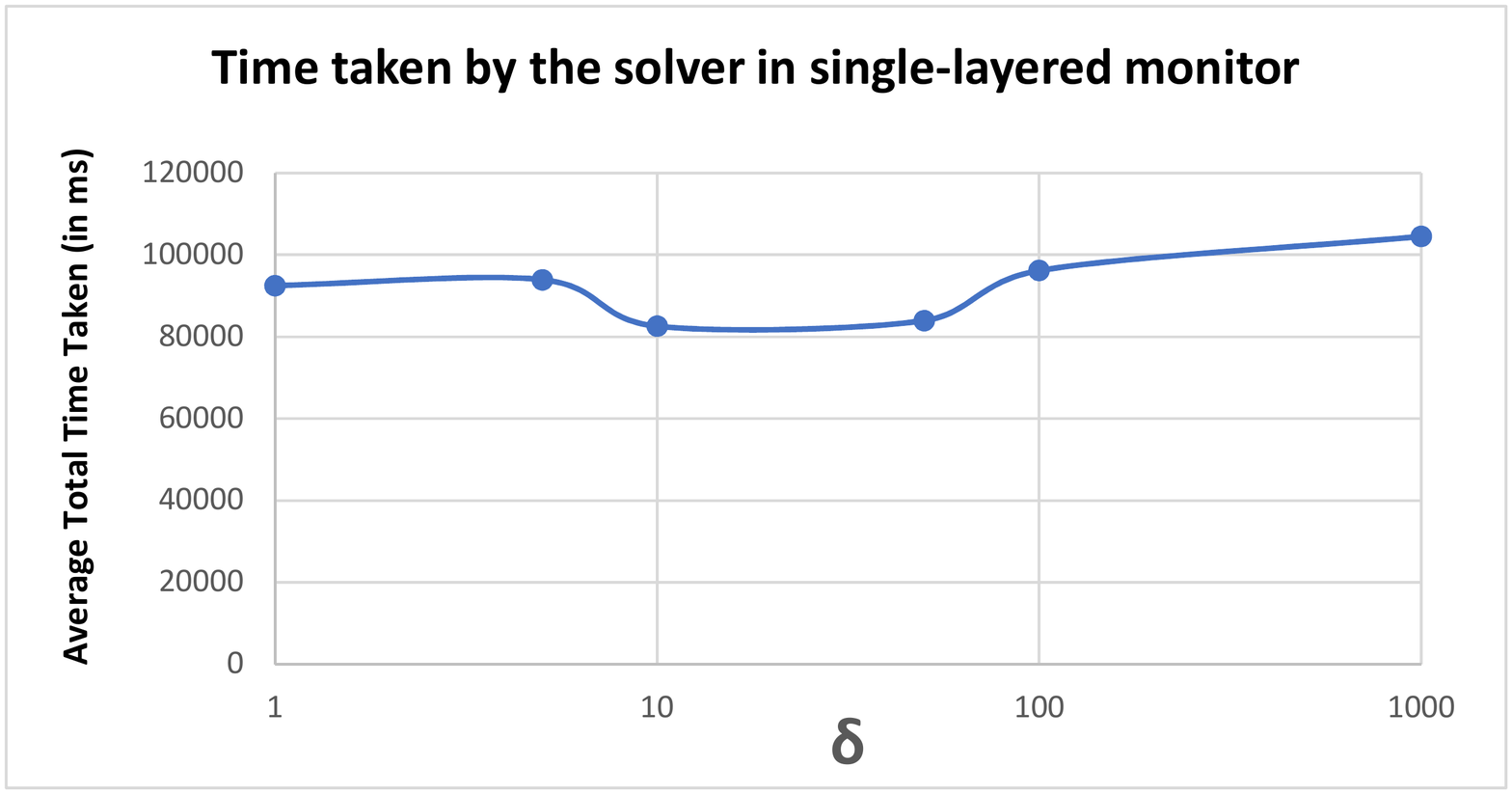}}
\subfloat[Varying $\delta$ and $\gamma$ (Time taken by the first layer was approximately 2 seconds)\label{fig:markwindgammadelta}]{\includegraphics[clip,trim=0.1cm 3.5cm 0.1cm 3.5cm,width=0.5\textwidth]{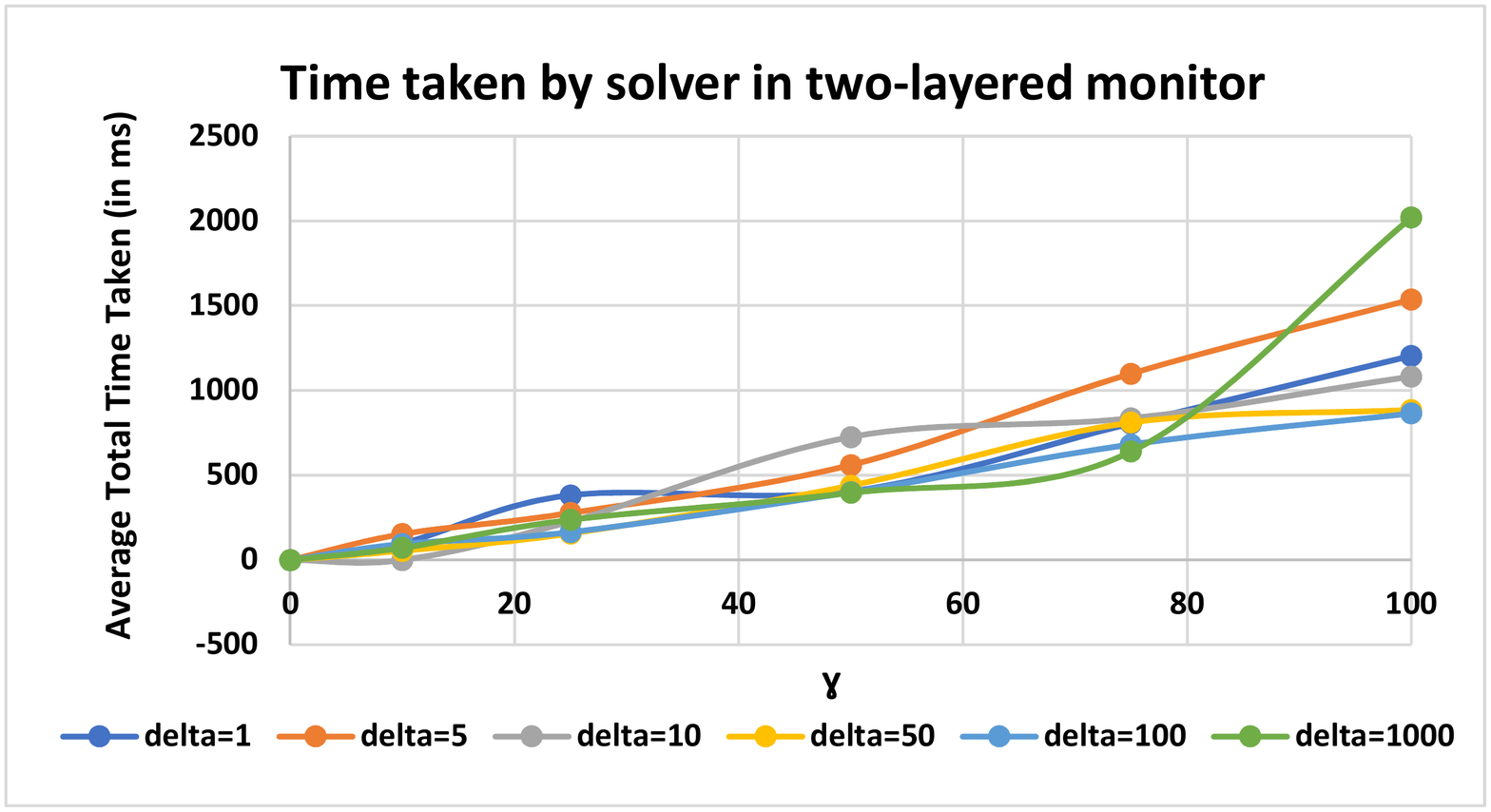}}
\\
\subfloat[Varying $\epsilon$ and $\gamma$\label{fig:allwindgammaeps}]{\includegraphics[clip,trim=0.1cm 3.5cm 0.1cm 3.5cm,width=0.5\textwidth]{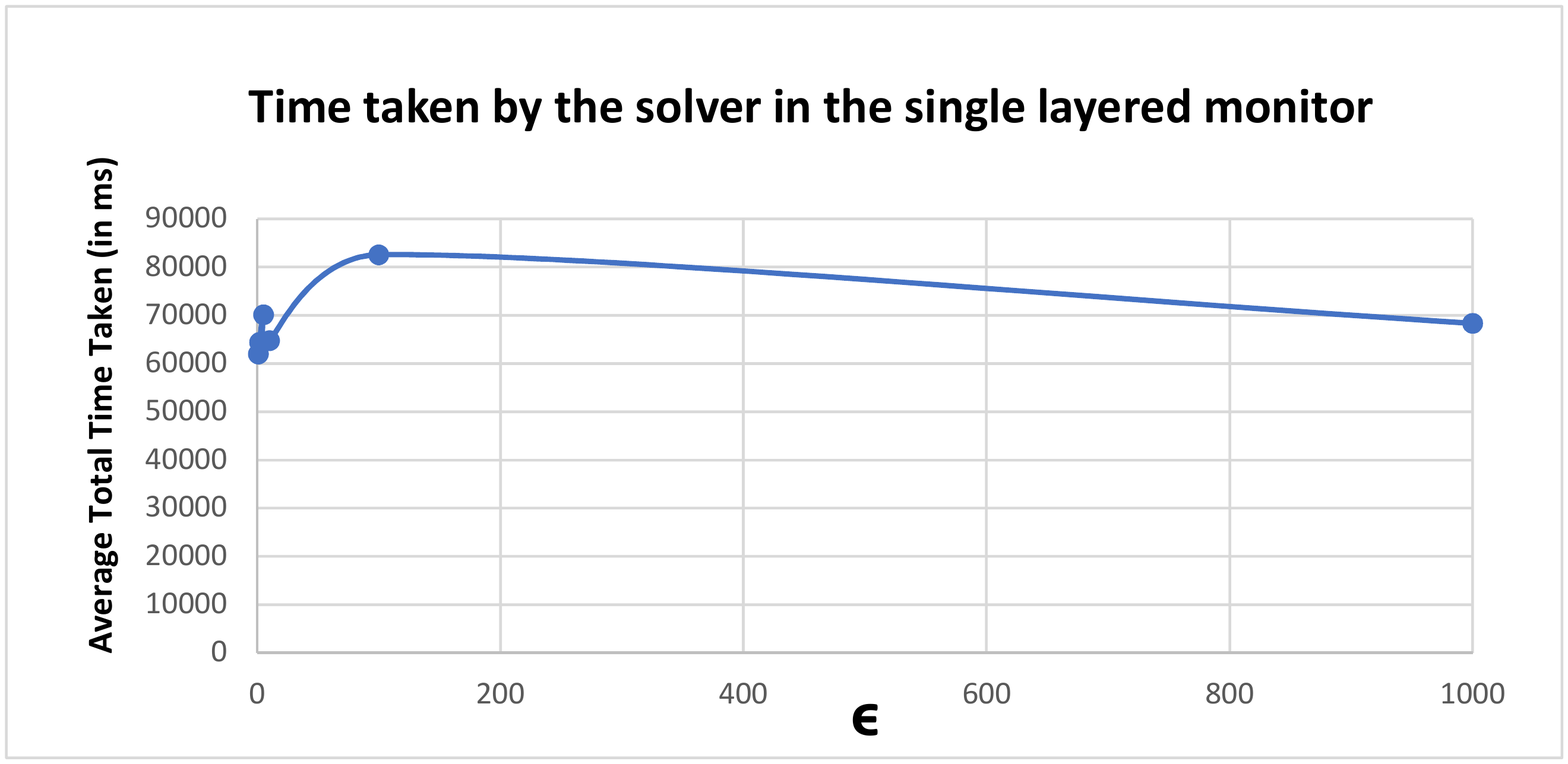}}
\subfloat[Varying $\epsilon$ and $\gamma$ (Time taken by the first layer was approximately 3 seconds)\label{fig:markwindgammaeps}]{\includegraphics[clip,trim=0.1cm 3.5cm 0.1cm 3.5cm,width=0.5\textwidth]{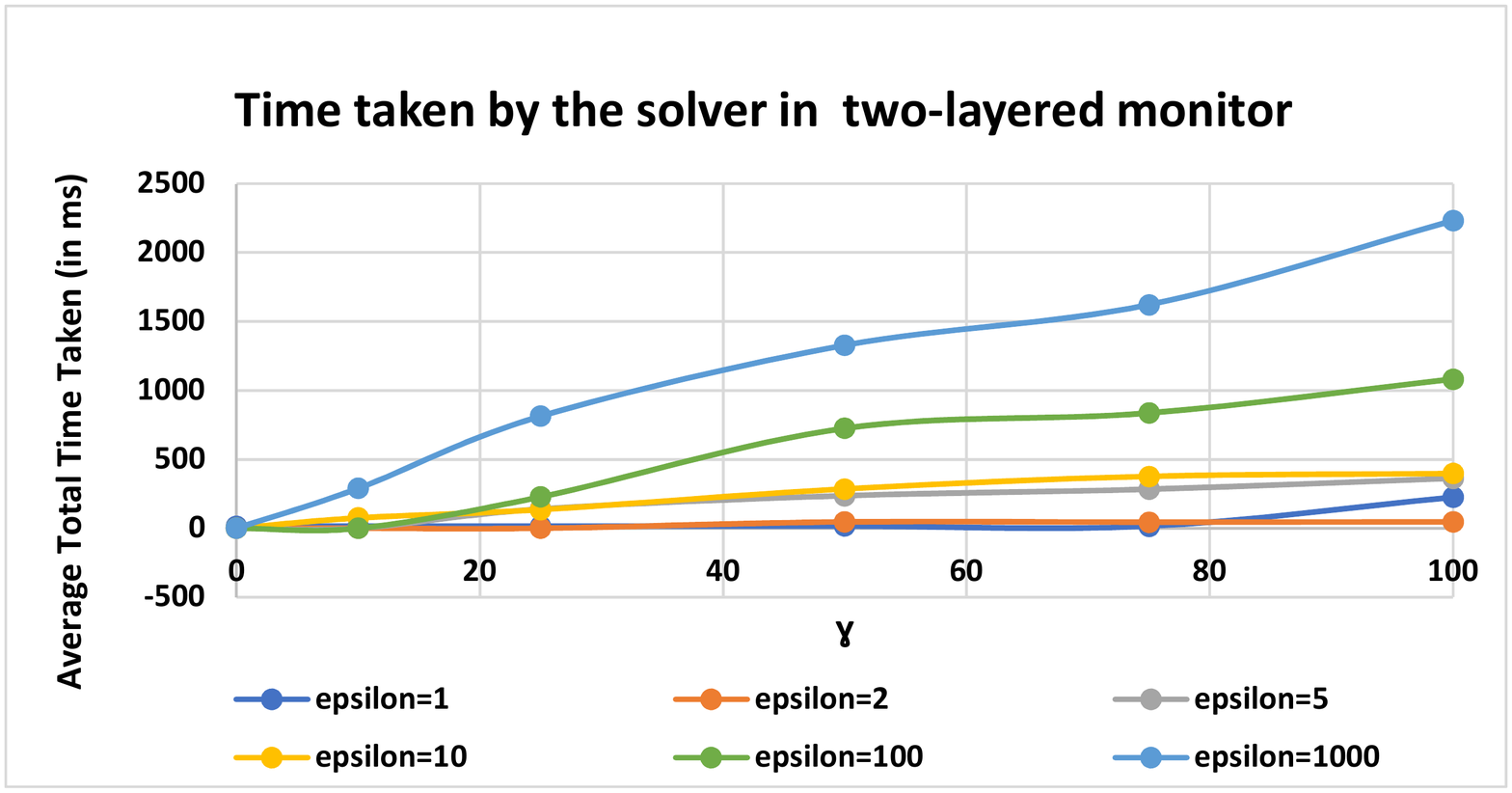}}
\\
\subfloat[Varying $\alpha$ and $\gamma$\label{fig:allwindgammaalpha}]{\includegraphics[clip,trim=0.1cm 3.5cm 0.1cm 3.5cm,width=0.5\textwidth]{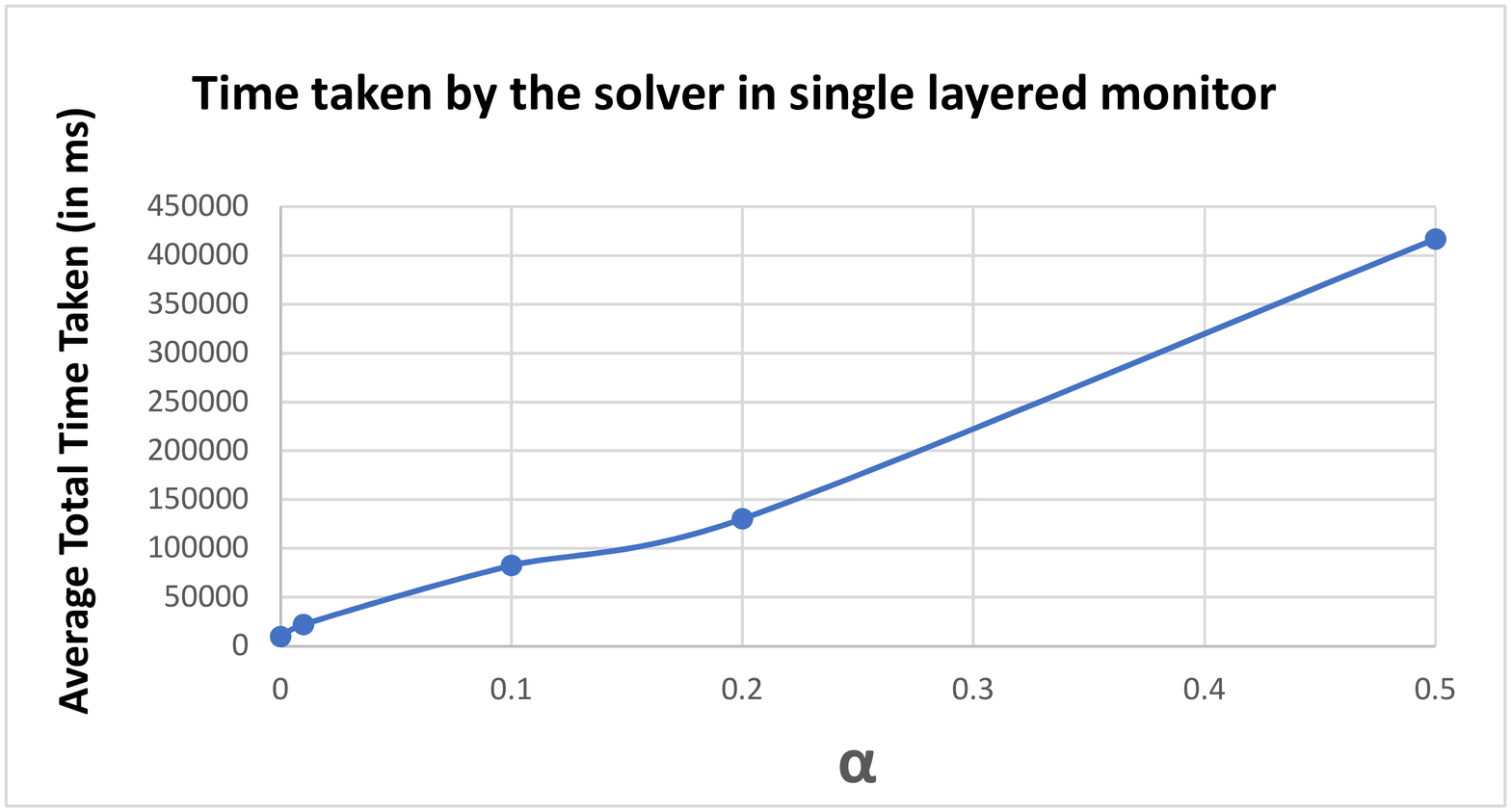}}
\subfloat[Varying $\alpha$ and $\gamma$ (Time taken by the first layer was approximately 3 seconds)\label{fig:markwindgammaalpha}]{\includegraphics[clip,trim=0.1cm 3.5cm 0.1cm 3.5cm,width=0.5\textwidth]{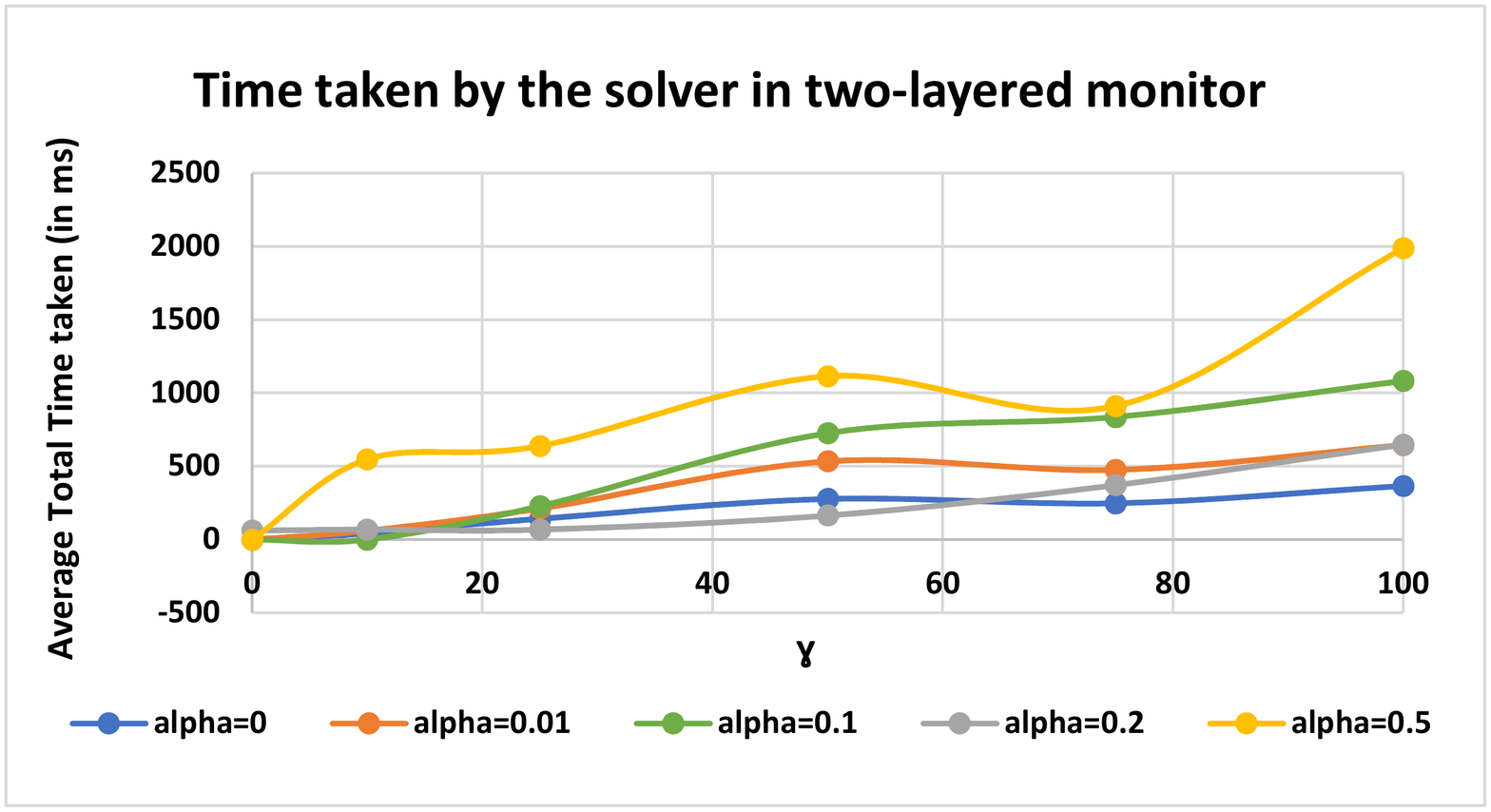}}
\caption{\small{Time taken by SMT Solver to detect violations of mutual exclusion in a time division multiplexing protocol when processing all windows vs windows marked by $\gamma$-extension. $\gamma$ is varied as fractions of $\epsilon$ (taking floor value if the fraction is not an integer). 
Time is measured in milliseconds. Default values: n=10, $\epsilon=100$, $\alpha=0.1$, $\delta=10$.}}
\label{fig:SolverTimeToken}
\vspace*{-6mm}
\end{figure*}

\section{Related Work}
\label{sec:relatedWork}

Predicate detection using different types of clocks have been studied extensively in the past. In \cite{Stoller2000}, Stoller proposed the use of physical clocks to detect global predicates, where for a system of $n$ processes, if the inter-event spacing $E$ is greater than the clock skew, then the total number of possible states to be evaluated is $O(En)$. 
Vector clocks are also widely used for predicate detection \cite{garg_weak,garg_conj_pred,ParallelPredDetGarg}. The disadvantage of using vector clocks is that each vector clock timestamp is of size $O(n)$ and they do not account for clock skew. Specifically, if the clock skew in the system is bounded by $\epsilon$, then two events that seem to be concurrent based on their vector clock timestamps are not concurrent events if they happened more than $\epsilon$ apart. So predicate detection using vector clocks can result in false positives.
One can use Optimal Vector Clocks \cite{optimalVC} to reduce the size of the timestamps or Hybrid Vector Clocks to reduce the size and account for clock skew. But, the size still remains $O(n)$ in the worst case. While Biased Hybrid Logical Clocks \cite{biasedClocks}, which are constant sized clocks, can be used to perform predicate detection, like HLC they suffer from false negatives. In this paper, we proposed the use of HLC, that are constant sized clocks, for predicate detection, and discussed how one can eliminate false negatives and/or false positives in it.

Several existing monitoring techniques use the notion of allowing false negatives for efficiency. Specifically, in \cite{optimisticHybridAnalysis}, they perform optimistic hybrid analysis by allowing false negatives during static analysis to speed up the dynamic analysis. However, to ensure the absence of false positives they occasionally require rollbacks and repeat the analysis without allowing false negatives during static analysis.
Monitoring approaches that use techniques like sampling 
aim to improve monitor efficiency by allowing false negatives. In \cite{LiteRace}, they improve efficiency by analyzing only selected portions of program executions. In \cite{RaceTrack}, they use adaptive tracking for efficient detection of data races. To eliminate false positives they switch the monitoring granularity from object level (for example tracking locksets associated with arrays) to field level (tracking locksets associated with array elements) only if the object level analysis reports a warning.

Monitors in \cite{adjustMonitorOverhead} and \cite{MonitoringWithDynamicPollingAndBuffer_Borzoo} adapt to overhead-budgets or timing/memory constraints. Specifically, in \cite{adjustMonitorOverhead} some monitoring operations are skipped to stay within predefined overhead budget while maximizing coverage (reduce false negatives) within the budget. In \cite{MonitoringWithDynamicPollingAndBuffer_Borzoo}, they proposed an approach where the monitor uses a feedback loop to control how frequently the events are reported to the monitor. This helps the monitor to adapt to the frequency of events in system. They also proposed the use of dynamically-sized buffers, where the size of the buffers that stored events between monitor invocations were changed to adapt to demand. Our two-layered monitoring approach can be combined with the approach in \cite{MonitoringWithDynamicPollingAndBuffer_Borzoo}. Specifically, our two-layered monitor has a fixed polling period (batches of fixed-length) and we consider unbounded buffers. We can extend our approach to let the polling periods and the buffer sizes to be dynamic. On the other hand, we could also extend our approach to use a feedback loop to learn about the frequency of events, and dynamically turn off the second layer of the monitor if the frequency is high at the cost of false positives/negatives (depending on $\gamma$), and turn it back on when the frequency is not too high. While the monitoring solutions provided in this paper do not dynamically adapt to timing/memory constraints or adjust monitoring operations to stay within predefined overhead-budgets, we presented different monitoring approaches that one can choose from to suit their monitoring budget and coverage.

\section{Conclusion}
\label{sec:conclusion}

In this paper, we focused on the problem of HLC based predicate detection (monitoring) with the use of $\gamma$-extension. A key advantage of this monitor is its simplicity and reduced overhead. Specifically, it permits the use of just $O(1)$ sized timestamps. Also, it permits monitoring of complex predicates. 

A key disadvantage of HLC based monitors was that they suffered from false negatives. We argued that in partially synchronous systems (that rely on clocks being synchronized within $\epsilon$), we can eliminate the false negatives at the cost of permitting false positives by using $\epsilon$-extension. By choosing a $\gamma$ extension ($\gamma < \epsilon$), we can obtain a trade-off between false positives and false negatives. 
We combined the HLC based monitor with $\gamma$-extension and the SMT based monitor in \cite{MonitoringUsingSMT} to obtain efficient monitors. When we set $\gamma =\epsilon$, these monitors have no false positives or false negatives (i.e., they identify all bugs without identifying any phantom bugs). Furthermore, the time required for monitors is significantly smaller (85-95\% less) than \cite{MonitoringUsingSMT}. 

In the event that this overhead is still higher than anticipated, $\gamma$-extension ($\gamma < \epsilon$) can help. Specifically, $\gamma$-extension reduces the number of false positives thereby reducing the instances when the SMT solver is called. The cost of this is that the monitor will suffer from false negatives. However,  if we are dealing with bugs that stay latent for a long time before they result in an actual problem, this may be acceptable when the other option is an inability to monitor. For example, for the application in \ref{sec:hlcgammaSMTExpeAnalysis}, when $\alpha=0.1,n=10,\delta=10,\epsilon=100$, if we choose $\gamma = 0.25*\epsilon$, then the cost of monitoring becomes 228 ms (when compared with 1.1s for $\epsilon$-extension). In this scenario, the false negative rate is 0.627, i.e., roughly every 3 bugs out of 8 would be identified.

We note that while our approach is designed for partially synchronous systems, it works for asynchronous ($\epsilon = \infty$) systems as well. Specifically, the notion of $\gamma$-extension can be applied for any $\gamma$, $\gamma < \epsilon$. The number of expected false negatives in this work can be characterized by the results in \cite{PrecRecallPartialSyncDistSysRV2016}. While this approach would have some false negatives, the overall cost of it would be significantly lower as it relies on $O(1)$ HLC timestamps than $O(n)$ vector clock timestamps.

\bibliographystyle{IEEEtran}
\bibliography{vidhya}

\clearpage
\appendices
\section{Detailed Analysis of Experimental Results for $\gamma$-extension}

In Tables \ref{tab:fpvConjPredInAppendix} and \ref{tab:fnvConjPredInAppendix}, we present the precision and recall results of a monitor that performs conjunctive predicate detection under different system parameters. 


\textbf{Precision} of the monitor is computed as the ratio of the number of \textit{valid} snapshots (i.e., those that are consistent and where $\pred$ is true)  detected by the monitor to the total number of snapshots reported by it. 
In the table we provide the actual ratio i.e. number of valid snapshots divided by the total number of snapshots in brackets, right next to the precision.
\textbf{Recall} of the monitor is computed as the ratio of the number of valid snapshots detected by the monitor to the number of actual valid snapshots in the system. 
In the table we provide the actual ratio i.e. number of valid snapshots detected divided by the number of actual snapshots in brackets, right next to the recall. 
We computed the precision and recall of the monitor under different system parameters namely $\beta$ - the rate at which the local predicate becomes true at a process, $\alpha$ - rate at which a process sends a message, $\epsilon$ - clock skew, $\delta$ - message delay, $\ell$ - duration for which the local predicate remains true at a process and $n$ - number of processes.

%
\textbf{Effect of $\beta$, $\ell$ and $n$.}
We observe that precision and recall increase with increase in $\beta$ and $\ell$, because as $\beta$ or $\ell$ increase they increase the probability of the local predicate being true in a consistent snapshot. We also observe that for a smaller $n$ (c.f. Tables \ref{tab:fpvBetaFiveProcInAppendix} and \ref{tab:fpvAlphaFiveProcInAppendix}) the monitor has a better precision and recall than for a higher $n$, because for an increased number of processes the probability of a conjunctive predicate ($\bigwedge\pred_i$) being true in a snapshot decreases.

\textbf{Effect of $\alpha$ and $\delta$.}
We observe that the precision decreases with increase in $\alpha$ and decrease in $\delta$. This observation is compatible with our discussion in Section \ref{sec:reduceFPVwithGamma}. Specifically, in Section \ref{sec:reduceFPVwithGamma}, we considered the scenario where a snapshot that is not a consistent snapshot due to communication (causality) between the events in the snapshot is incorrectly declared as a consistent snapshot by the monitor using $\gamma$-extension. As $\alpha$ increases - communication increases and as $\delta$ decreases - messages are not delayed (causality has immediate effect), therefore the false positives increase and precision of the monitor decreases.
On the other hand, recall increases with increase in $\alpha$ and decrease in $\delta$. This is because in general an increase in $\alpha$ and decrease in $\delta$ reduces the chance of a snapshot being a consistent snapshot, thereby reducing the number of actual valid snapshots in the system. A decrease in the number of actual valid snapshots in the system increases the recall.

\textbf{Effect of $\gamma$.}  We observe that irrespective of the underlying system setting the precision of the monitor decreases and the recall increases as the value of $\gamma$ increases. For example in  Table \ref{tab:fpvBetaInAppendix}, consider the case where $\beta=0.045$, as the value of $\gamma$ increases from $0.1 * \epsilon$ to $\epsilon$, the precision of the monitor drops from $0.929$ to $0.295$.\footnote{Note that the scenarios where the precision is NA even when the value of $\gamma$ is small (for example $\beta=0.02, \gamma=0.1* \epsilon$) corresponds to the case where the monitor does not detect any snapshots and therefore detects no valid snapshots. In other words, these scenarios correspond to the case where the monitor reports no false positives, therefore the precision is implicitly perfect.}
On the other hand, in Table \ref{tab:fnvBetaInAppendix}, for the case $\beta=0.045$, as the value of $\gamma$ increases from $0.1 * \epsilon$ to $\epsilon$ the recall of the monitor increases from $0.009$ to $1$.
Thus if $\gamma$ is set to be a smaller fraction of $\epsilon$ then the monitor will have higher precision and lower recall. On the other hand if $\gamma$ is set to be closer to $\epsilon$ then the monitor will have lower precision and higher recall. Observe that one can choose $\gamma$ such that both precision and recall of the monitor are reasonable and not too low. For example, in Table \ref{tab:fpvBetaFiveProcInAppendix} for $\beta=0.02$ when $\gamma$ is set to $0.5*\epsilon$ the precision is $0.751$ and the corresponding recall in Table \ref{tab:fnvBetaFiveProcInAppendix} is $0.824$.
%

While the observed precision values are low, observe that the number of actual valid snapshots in the system for some of the settings are very high. For example, in Table \ref{tab:fpvBetaInAppendix} for $\beta=0.045$, $\gamma=\epsilon$, precision is $0.295$, and the number of actual valid snapshots in the system is $2907$ (c.f. Table \ref{tab:fnvBetaInAppendix}, $\beta=0.045$, $\gamma=\epsilon$, denominator). This corresponds to the case where the monitor has a low precision in a system that has several bugs. In such a scenario, bugs in the system could be related to a common problem and detecting a few bugs and fixing the cause could eliminate majority of the bugs in the system. Also, on the other hand if a system has so many bugs one should be able to detect them by generic testing.
\begin{table*}[ht]
\centering{\textbf{No. of Valid Snapshots/Total Snapshots Detected (Default values: n = 10, $\epsilon$ = 100, $\alpha$ = 0.1, $\delta$ = 10, $\ell$ = 1, $\beta$ = 0.02)}}\\
\subfloat[Varying $\beta$ - rate at which the local predicate becomes true at a process]{\label{tab:fpvBetaInAppendix}
\begin{tabular}{|l|l|l|l|l|l|l|}
\toprule
\textbf{Precision} & \multicolumn{1}{l|}{\textbf{$\beta$ = 0.02}} & \multicolumn{1}{l|}{\textbf{$\beta$ = 0.025}} & \multicolumn{1}{l|}{\textbf{$\beta$ = 0.03}} & \multicolumn{1}{l|}{\textbf{$\beta$ = 0.035}} & \multicolumn{1}{l|}{\textbf{$\beta$ = 0.04}} & \multicolumn{1}{l|}{\textbf{$\beta$ = 0.045}} \\
\midrule
\textbf{$\gamma$ = 0.10 * $ \epsilon$} & NA & NA & 0.333 (1/3) & 1.000 (4/4) & 0.739 (17/23) & 0.929 (26/28)\\
\textbf{$\gamma$ = 0.15 * $ \epsilon$} & 0.500 (1/2) & 0.000 (0/3) & 0.529 (9/17)& 0.650 (13/20)& 0.797 (102/128)& 0.743 (113/152)\\
\textbf{$\gamma$ = 0.20 * $ \epsilon$} & 0.500 (2/4)& 0.286 (2/7)& 0.511 (46/90)& 0.582 (46/79)& 0.732 (341/466)& 0.742 (339/457)\\
\textbf{$\gamma$ = 0.25 * $ \epsilon$} & 0.200 (3/15)& 0.190 (4/21)& 0.477 (126/264)& 0.466 (110/236)& 0.688 (778/1131)& 0.706 (771/1092)\\
\textbf{$\gamma$ = 0.50 * $ \epsilon$} & 0.077 (44/572)& 0.070 (42/604)& 0.192 (699/3635)& 0.192 (722/3756)& 0.383 (2756/7188)& 0.384 (2780/7246)\\
\textbf{$\gamma$ = 0.75  * $ \epsilon$} & 0.025 (71/2827)& 0.024 (70/2931)& 0.107 (809/7575)& 0.109 (844/7721)& 0.304 (2877/9473)& 0.305 (2905/9518)\\
\textbf{$\gamma$ = $\epsilon$} & 0.013 (71/5651)& 0.013 (71/5665)& 0.088 (813/9208)& 0.092 (850/9252)& 0.292 (2879/9855)& 0.295 (2907/9859)\\
\hline
\end{tabular}}
\qquad
\subfloat[Varying $\alpha$ - rate at which processes send messages]{\label{tab:fpvAlphaInAppendix}
\begin{tabular}{|l|l|l|l|}
\toprule
\textbf{Precision} & \multicolumn{1}{l|}{\textbf{$\alpha$ = 0.01}} & \multicolumn{1}{l|}{\textbf{$\alpha$ = 0.1}} & \multicolumn{1}{l|}{\textbf{$\alpha$ = 0.2}} \\
\midrule
\textbf{$\gamma$ = 0.10 * $ \epsilon$} & 1.000 (9/9)& NA& NA\\
\textbf{$\gamma$ = 0.15 * $ \epsilon$} & 0.913 (21/23)& 0.500 (1/2)& NA\\
\textbf{$\gamma$ = 0.20 * $ \epsilon$} & 0.911 (51/56)& 0.500 (2/4)& NA\\
\textbf{$\gamma$ = 0.25 * $ \epsilon$} & 0.901 (136/151)& 0.200 (3/15)& 0.000 (0/1)\\
\textbf{$\gamma$ = 0.50 * $ \epsilon$} & 0.830 (1373/1655)& 0.077 (44/572)& 0.006 (1/164)\\
\textbf{$\gamma$ = 0.75  * $ \epsilon$} & 0.688 (3348/4867)& 0.025 (71/2827)& 0.001 (1/1131)\\
\textbf{$\gamma$ = $\epsilon$} & 0.541 (4141/7657)& 0.013 (71/5651)& 0.000 (1/3084)\\
\hline
\end{tabular}}
\qquad
\subfloat[Varying $\epsilon$-clock skew]{\label{tab:fpvEpsilonInAppendix}
\begin{tabular}{|l|l|l|}
\toprule
\textbf{Precision} & \multicolumn{1}{l|}{\textbf{$\epsilon$ = 100}} & \multicolumn{1}{l|}{\textbf{$\epsilon$ = 1000}} \\
\midrule
\textbf{$\gamma$ = 0.10 * $ \epsilon$} & NA& 0.077 (76/993)\\
\textbf{$\gamma$ = 0.15 * $ \epsilon$} & 0.500 (1/2)& 0.076 (76/1000)\\
\textbf{$\gamma$ = 0.20 * $ \epsilon$} & 0.500 (2/4)& 0.076 (76/1000)\\
\textbf{$\gamma$ = 0.25 * $ \epsilon$} & 0.200 (3/5)& 0.076 (76/1000)\\
\textbf{$\gamma$ = 0.50 * $ \epsilon$} & 0.077 (44/572)& 0.076 (76/1000)\\
\textbf{$\gamma$ = 0.75  * $ \epsilon$} & 0.025 (71/2827)& 0.076 (76/1000)\\
\textbf{$\gamma$ = $\epsilon$} & 0.013 (71/5651)& 0.076 (76/999)\\
\hline
\end{tabular}}
\qquad
\subfloat[Varying $\delta$ - message delay]{\label{tab:fpvDeltaInAppendix}\begin{tabular}{|l|l|l|l|l|l|l|}
\toprule
\textbf{Precision} & \multicolumn{1}{l|}{\textbf{$\delta$ = 1}} & \multicolumn{1}{l|}{\textbf{$\delta$ = 5}} & \multicolumn{1}{l|}{\textbf{$\delta$ = 10}} & \multicolumn{1}{l|}{\textbf{$\delta$ = 50}} & \multicolumn{1}{l|}{\textbf{$\delta$ = 100}} & \multicolumn{1}{l|}{\textbf{$\delta$ = 1000}} \\
\midrule
\textbf{$\gamma$ = 0.10 * $ \epsilon$} & NA& NA& NA& NA& NA& NA\\
\textbf{$\gamma$ = 0.15 * $ \epsilon$} & 0.000 (0/5)& 0.500 (1/2)& 0.500 (1/2)& 1.000 (1/1)& NA& 1.000 (1/1)\\
\textbf{$\gamma$ = 0.20 * $ \epsilon$} & 0.000 (0/10)& 0.286 (2/7)& 0.500 (2/4)& 1.000 (5/5)& NA& 1.000 (4/4)\\
\textbf{$\gamma$ = 0.25 * $ \epsilon$} & 0.054 (2/37)& 0.111 (3/27)& 0.200 (3/15)& 0.800 (8/10)& 0.909 (10/11)& 1.000 (8/8)\\
\textbf{$\gamma$ = 0.50 * $ \epsilon$} & 0.010 (7/678)& 0.042 (25/602)& 0.077 (44/572)& 0.752 (415/552)& 0.846 (452/534)& 0.871 (452/519)\\
\textbf{$\gamma$ = 0.75  * $ \epsilon$} & 0.002 (7/2911)& 0.011 (31/2864)& 0.025 (71/2827)& 0.565 (1535/2717)& 0.800 (2176/2719)& 0.794 (2144/2701)\\
\textbf{$\gamma$ = $\epsilon$} & 0.001 (8/5768)& 0.006 (31/5578)& 0.013 (71/5651)& 0.382 (2116/5536)& 0.679 (3737/5502)& 0.693 (3759/5426)\\
\hline
\end{tabular}}
\qquad
\subfloat[Varying $\ell$- duration for which the local predicate remains true]{\label{tab:fpvIntervalLengthInAppendix}\begin{tabular}{|l|l|l|l|l|l|l|}
\toprule
\textbf{Precision} & \multicolumn{1}{l|}{\textbf{$\ell$ = 1}} & \multicolumn{1}{l|}{\textbf{$\ell$ = 10}} & \multicolumn{1}{l|}{\textbf{$\ell$ = 20}} & \multicolumn{1}{l|}{\textbf{$\ell$ = 50}} & \multicolumn{1}{l|}{\textbf{$\ell$ = 100}} & \multicolumn{1}{l|}{\textbf{$\ell$ = 1000}} \\
\midrule
\textbf{$\gamma$ = 0.10 * $ \epsilon$} & NA& 0.500 (1/2)& 0.923 (12/13)& 0.859 (140/163)& 0.887 (524/591)& 0.908 (1337/1472)\\
\textbf{$\gamma$ = 0.15 * $ \epsilon$} & 0.500 (1/2)& 0.600 (6/10)& 0.765 (26/34)& 0.801 (233/291)& 0.860 (762/886)& 0.892 (1653/1854)\\
\textbf{$\gamma$ = 0.20 * $ \epsilon$} & 0.500 (2/4)& 0.576 (19/33)& 0.675 (54/80)& 0.756 (356/471)& 0.829 (1035/1249)& 0.867 (1984/2289)\\
\textbf{$\gamma$ = 0.25 * $ \epsilon$} & 0.200 (3/15)& 0.476 (40/84)& 0.570 (102/179)& 0.704 (533/757)& 0.776 (1295/1669)& 0.832 (2309/2775)\\
\textbf{$\gamma$ = 0.50 * $ \epsilon$} & 0.077 (44/572)& 0.183 (191/1041)& 0.275 (434/1581)& 0.421 (1325/3150)& 0.546 (2419/4430)& 0.629 (3451/5483)\\
\textbf{$\gamma$ = 0.75  * $ \epsilon$} & 0.025 (71/2827)& 0.070 (261/3744)& 0.126 (551/4373)& 0.262 (1555/5934)& 0.388 (2664/6868)& 0.492 (3701/7525)\\
\textbf{$\gamma$ = $\epsilon$} & 0.013 (71/5651)& 0.041 (264/6368)& 0.082 (566/6898)& 0.201 (1568/7812)& 0.320 (2687/8394)& 0.431 (3723/8639)\\
\hline
\end{tabular}}
\\
\vspace{7pt}
\centering{\textbf{No. of Valid Snapshots/Total Snapshots Detected (Default values: n = 5, $\epsilon$ = 100, $\alpha$ = 0.1, $\delta$ = 10, $\ell$ = 20,  $\beta$ = 0.004)}}\\
\subfloat[Varying $\beta$ - rate at which the local predicate becomes true at a process]{\label{tab:fpvBetaFiveProcInAppendix}\begin{tabular}{|l|l|l|l|l|l|}
\toprule
\textbf{Precision} & \multicolumn{1}{l|}{\textbf{$\beta$ = 0.02}} & \multicolumn{1}{l|}{\textbf{$\beta$ = 0.01}} & \multicolumn{1}{l|}{\textbf{$\beta$ = 0.005}} & \multicolumn{1}{l|}{\textbf{$\beta$ = 0.004}} & \multicolumn{1}{l|}{\textbf{$\beta$ = 0.002}} \\
\midrule
\textbf{$\gamma$ = 0.10 * $ \epsilon$} & 0.953 (784/823)& 0.915 (483/528)& 0.700 (7/10)& 1.000 (2/2)& NA\\
\textbf{$\gamma$ = 0.15 * $ \epsilon$} & 0.937 (1145/1222)& 0.904 (768/850)& 0.667 (12/18)& 0.667 (2/3)& NA\\
\textbf{$\gamma$ = 0.20 * $ \epsilon$} & 0.913 (1548/1696)& 0.869 (1059/1218)& 0.710 (22/31)& 0.667 (2/3)& NA\\
\textbf{$\gamma$ = 0.25 * $ \epsilon$} & 0.893 (1977/2214)& 0.847 (1409/1663)& 0.596 (28/47)& 0.600 (3/5)& NA\\
\textbf{$\gamma$ = 0.50 * $ \epsilon$} & 0.751 (3856/5133)& 0.666 (3011/4518)& 0.376 (86/229)& 0.286 (12/42)& 0.000 (0/2)\\
\textbf{$\gamma$ = 0.75  * $ \epsilon$} & 0.628 (4532/7222)& 0.535 (3634/6789)& 0.182 (141/776)& 0.135 (26/192)& 0.333 (1/3)\\
\textbf{$\gamma$ = $\epsilon$} & 0.556 (4681/8426)& 0.462 (3756/8123)& 0.104 (184/1766)& 0.067 (38/566)& 0.071 (1/14)\\
\hline
\end{tabular}}
\qquad
\subfloat[Varying $\alpha$ - rate at which processes send messages]{\label{tab:fpvAlphaFiveProcInAppendix}\begin{tabular}{|l|l|l|l|}
\toprule
\textbf{Precision} & \multicolumn{1}{l|}{\textbf{$\alpha$ = 0.01}} & \multicolumn{1}{l|}{\textbf{$\alpha$ = 0.025}} & \multicolumn{1}{l|}{\textbf{$\alpha$ = 0.05}} \\
\midrule
\textbf{$\gamma$ = 0.10 * $ \epsilon$} & 1.000 (2/2)& 1.000 (2/2)& 0.667 (2/3)\\
\textbf{$\gamma$ = 0.15 * $ \epsilon$} & 1.000 (3/3)& 1.000 (3/3)& 0.667 (2/3)\\
\textbf{$\gamma$ = 0.20 * $ \epsilon$} & 1.000 (3/3)& 1.000 (3/3)& 0.667 (2/3)\\
\textbf{$\gamma$ = 0.25 * $ \epsilon$} & 1.000 (6/6)& 0.833 (5/6)& 0.667 (4/6)\\
\textbf{$\gamma$ = 0.50 * $ \epsilon$} & 0.850 (34/40)& 0.590 (23/39)& 0.390 (16/41)\\
\textbf{$\gamma$ = 0.75  * $ \epsilon$} & 0.631 (113/179)& 0.425 (74/174)& 0.271 (49/181)\\
\textbf{$\gamma$ = $\epsilon$} & 0.493 (276/560)& 0.310 (174/562)& 0.157 (90/572)\\
\hline
\end{tabular}}
\qquad%
\caption{Percentage of Valid Snapshots out of all snapshots detected during Conjunctive Predicate Detection using $\gamma$-extension.\label{tab:fpvConjPredInAppendix}}
\end{table*}%


\begin{table*}
\centering{\textbf{No. of Valid Snapshots detected/Total No. of Valid Snapshots in the system(Default values: n = 10, $\epsilon$ = 100, $\alpha$ = 0.1, $\delta$ = 10, $\ell$ = 1, $\beta$ = 0.02)}}\\
\subfloat[Varying $\beta$ - rate at which the local predicate becomes true at a process]{\label{tab:fnvBetaInAppendix}\begin{tabular}{|l|l|l|l|l|l|l|}
\toprule
\textbf{Recall} & \multicolumn{1}{l|}{\textbf{$\beta$ = 0.02}} & \multicolumn{1}{l|}{\textbf{$\beta$ = 0.025}} & \multicolumn{1}{l|}{\textbf{$\beta$ = 0.03}} & \multicolumn{1}{l|}{\textbf{$\beta$ = 0.035}} & \multicolumn{1}{l|}{\textbf{$\beta$ = 0.04}} & \multicolumn{1}{l|}{\textbf{$\beta$ = 0.045}} \\
\midrule
\textbf{$\gamma$ = 0.10 * $ \epsilon$} & 0.000 (0/71)& 0.000 (0/71)& 0.001 (1/813)& 0.005 (4/850)& 0.006 (17/2879)& 0.009 (26/2907)\\
\textbf{$\gamma$ = 0.15 * $ \epsilon$} & 0.014 (1/71)& 0.000 (0/71)& 0.011 (9/813)& 0.015 (13/850)& 0.035 (102/2879)& 0.039 (113/2907)\\
\textbf{$\gamma$ = 0.20 * $ \epsilon$} & 0.028 (2/71)& 0.028 (2/71)& 0.057 (46/813)& 0.054 (46/850)& 0.118 (341/2879)& 0.117 (339/2907)\\
\textbf{$\gamma$ = 0.25 * $ \epsilon$} & 0.042 (3/71)& 0.056 (4/71)& 0.155 (126/813)& 0.129 (110/850)& 0.270 (778/2879)& 0.265 (771/2907)\\
\textbf{$\gamma$ = 0.50 * $ \epsilon$} & 0.620 (44/71)& 0.592 (42/71)& 0.860 (699/813)& 0.849 (722/850)& 0.957 (2756/2879)& 0.956 (2780/2907)\\
\textbf{$\gamma$ = 0.75  * $ \epsilon$} & 1.000 (71/71)& 0.986 (70/71)& 0.995 (809/813)& 0.993 (844/850)& 0.999 (2877/2879)& 0.999 (2905/2907)\\
\textbf{$\gamma$ = $\epsilon$} & 1.000 (71/71)& 1.000 (71/71)& 1.000 (813/813)& 1.000 (850/850)& 1.000 (2879/2879)& 1.000 (2907/2907)\\
\hline
\end{tabular}}
\qquad
\subfloat[Varying $\alpha$ - rate at which processes send messages]{\label{tab:fnvAlphaInAppendix}\begin{tabular}{|l|l|l|l|}
\toprule
\textbf{Recall} & \multicolumn{1}{l|}{\textbf{$\alpha$ = 0.01}} & \multicolumn{1}{l|}{\textbf{$\alpha$ = 0.1}} & \multicolumn{1}{l|}{\textbf{$\alpha$ = 0.2}} \\
\midrule
\textbf{$\gamma$ = 0.10 * $ \epsilon$} & 0.002 (9/4141)& 0.000 (0/71)& 0.000 (0/1)\\
\textbf{$\gamma$ = 0.15 * $ \epsilon$} & 0.005 (21/4141)& 0.014 (1/71)& 0.000 (0/1)\\
\textbf{$\gamma$ = 0.20 * $ \epsilon$} & 0.012 (51/4141)& 0.028 (2/71)& 0.000 (0/1)\\
\textbf{$\gamma$ = 0.25 * $ \epsilon$} & 0.033 (136/4141)& 0.042 (3/71)& 0.000 (0/1)\\
\textbf{$\gamma$ = 0.50 * $ \epsilon$} & 0.332 (1373/4141)& 0.620 (44/71)& 1.000 (1/1)\\
\textbf{$\gamma$ = 0.75  * $ \epsilon$} & 0.809 (3348/4141)& 1.000 (71/71)& 1.000 (1/1)\\
\textbf{$\gamma$ = $\epsilon$} & 1.000 (4141/4141)& 1.000 (71/71)& 1.000 (1/1)\\
\hline
\end{tabular}}
\qquad
\subfloat[Varying $\epsilon$-clock skew]{\label{tab:fnvEpsilonInAppendix}\begin{tabular}{|l|l|l|}
\toprule
\textbf{Recall} & \multicolumn{1}{l|}{\textbf{$\epsilon$ = 100}} & \multicolumn{1}{l|}{\textbf{$\epsilon$ = 1000}} \\
\midrule
\textbf{$\gamma$ = 0.10 * $ \epsilon$} & 0.000 (0/71)& 1.000 (76/76)\\
\textbf{$\gamma$ = 0.15 * $ \epsilon$} & 0.014 (1/71)& 1.000 (76/76)\\
\textbf{$\gamma$ = 0.20 * $ \epsilon$} & 0.028 (2/71)& 1.000 (76/76)\\
\textbf{$\gamma$ = 0.25 * $ \epsilon$} & 0.042 (3/71)& 1.000 (76/76)\\
\textbf{$\gamma$ = 0.50 * $ \epsilon$} & 0.620 (44/71)& 1.000 (76/76)\\
\textbf{$\gamma$ = 0.75  * $ \epsilon$} & 1.000 (71/71)& 1.000 (76/76)\\
\textbf{$\gamma$ = $\epsilon$} & 1.000 (71/71)& 1.000 (76/76)\\
\hline
\end{tabular}}
\qquad
\subfloat[Varying $\delta$ - message delay]{\label{tab:fnvDeltaInAppendix}\begin{tabular}{|l|l|l|l|l|l|l|}
\toprule
\textbf{Recall} & \multicolumn{1}{l|}{\textbf{$\delta$ = 1}} & \multicolumn{1}{l|}{\textbf{$\delta$ = 5}} & \multicolumn{1}{l|}{\textbf{$\delta$ = 10}} & \multicolumn{1}{l|}{\textbf{$\delta$ = 50}} & \multicolumn{1}{l|}{\textbf{$\delta$ = 100}} & \multicolumn{1}{l|}{\textbf{$\delta$ = 1000}} \\
\midrule
\textbf{$\gamma$ = 0.10 * $ \epsilon$} & 0.000 (0/8)& 0.000 (0/31)& 0.000 (0/71)& 0.000 (0/2116)& 0.000 (0/3737)& 0.000 (0/3759)\\
\textbf{$\gamma$ = 0.15 * $ \epsilon$} & 0.000 (0/8)& 0.032 (1/31)& 0.014 (1/71)& 0.000 (1/2116)& 0.000 (0/3737)& 0.000 (1/3759)\\
\textbf{$\gamma$ = 0.20 * $ \epsilon$} & 0.000 (0/8)& 0.065 (2/31)& 0.028 (2/71)& 0.002 (5/2116)& 0.000 (0/3737)& 0.001 (4/3759)\\
\textbf{$\gamma$ = 0.25 * $ \epsilon$} & 0.250 (2/8)& 0.097 (3/31)& 0.042 (3/71)& 0.004 (8/2116)& 0.003 (10/3737)& 0.002 (8/2759)\\
\textbf{$\gamma$ = 0.50 * $ \epsilon$} & 0.875 (7/8)& 0.806 (25/31)& 0.620 (44/71)& 0.196 (415/2116)& 0.121 (452/3737)& 0.120 (452/3759)\\
\textbf{$\gamma$ = 0.75  * $ \epsilon$} & 0.875 (7/8)& 1.000 (31/31)& 1.000 (71/71)& 0.725 (1535/2116)& 0.582 (2176/3737)& 0.570 (2144/3759)\\
\textbf{$\gamma$ = $\epsilon$} & 1.000 (8/8)& 1.000 (31/31)& 1.000 (71/71)& 1.000 (2116/2116)& 1.000 (3737/3737)& 1.000 (3759/3759)\\
\hline
\end{tabular}}
\qquad
\subfloat[Varying $\ell$- duration for which the local predicate remains true]{\label{tab:fnvIntervalLengthInAppendix}\begin{tabular}{|l|l|l|l|l|l|l|}
\toprule
\textbf{Recall} & \multicolumn{1}{l|}{\textbf{$\ell$ = 1}} & \multicolumn{1}{l|}{\textbf{$\ell$ = 10}} & \multicolumn{1}{l|}{\textbf{$\ell$ = 20}} & \multicolumn{1}{l|}{\textbf{$\ell$ = 50}} & \multicolumn{1}{l|}{\textbf{$\ell$ = 100}} & \multicolumn{1}{l|}{\textbf{$\ell$ = 1000}} \\
\midrule
\textbf{$\gamma$ = 0.10 * $ \epsilon$} & 0.000 (0/71)& 0.004 (1/264)& 0.021 (12/566)& 0.089 (140/1568)& 0.195 (524/2687)& 0.359 (1337/3723)\\
\textbf{$\gamma$ = 0.15 * $ \epsilon$} & 0.014 (1/71)& 0.023 (6/264)& 0.046 (26/566)& 0.149 (233/1568)& 0.284 (762/2687)& 0.444 (1653/3723)\\
\textbf{$\gamma$ = 0.20 * $ \epsilon$} & 0.028 (2/71)& 0.072 (19/264)& 0.095 (54/566)& 0.227 (356/1568)& 0.385 (1035/2687)& 0.533 (1984/3723)\\
\textbf{$\gamma$ = 0.25 * $ \epsilon$} & 0.042 (3/71)& 0.152 (40/264)& 0.180 (102/566)& 0.340 (533/1568)& 0.482 (1295/2687)& 0.620 (2309/3723)\\
\textbf{$\gamma$ = 0.50 * $ \epsilon$} & 0.620 (44/71)& 0.723 (191/264)& 0.767 (434/566)& 0.845 (1325/1568)& 0.900 (2419/2687)& 0.927 (3451/3723)\\
\textbf{$\gamma$ = 0.75  * $ \epsilon$} & 1.000 (71/71)& 0.989 (261/264)& 0.973 (551/566)& 0.992 (1555/1568)& 0.991 (2664/2687)& 0.994 (3701/3723)\\
\textbf{$\gamma$ = $\epsilon$} & 1.000 (71/71)& 1.000 (264/264)& 1.000 (566/566)& 1.000 (1568/1568)& 1.000 (2687/2687)& 1.000 (3723/3723)\\
\hline
\end{tabular}}
\\
\vspace{7pt}
\centering{\textbf{No. of Valid Snapshots detected/Total No. of Valid Snapshots in the system (Default values: n = 5, $\epsilon$ = 100, $\alpha$ = 0.1, $\delta$ = 10, $\ell$ = 20,  $\beta$ = 0.004)}}\\
\subfloat[Varying $\beta$ - rate at which the local predicate becomes true at a process]{\label{tab:fnvBetaFiveProcInAppendix}\begin{tabular}{|l|l|l|l|l|l|}
\toprule
\textbf{Recall} & \multicolumn{1}{l|}{\textbf{$\beta$ = 0.02}} & \multicolumn{1}{l|}{\textbf{$\beta$ = 0.01}} & \multicolumn{1}{l|}{\textbf{$\beta$ = 0.005}} & \multicolumn{1}{l|}{\textbf{$\beta$ = 0.004}} & \multicolumn{1}{l|}{\textbf{$\beta$ = 0.002}} \\
\midrule
\textbf{$\gamma$ = 0.10 * $ \epsilon$} & 0.167 (784/4681)& 0.129 (483/3756)& 0.038 (7/184)& 0.053 (2/38)& NA\\
\textbf{$\gamma$ = 0.15 * $ \epsilon$} & 0.245 (1145/4681)& 0.204 (768/3756)& 0.065 (12/184)& 0.053 (2/38)& NA\\
\textbf{$\gamma$ = 0.20 * $ \epsilon$} & 0.331 (1548/4681)& 0.282 (1059/3756)& 0.120 (22/184)& 0.053 (2/38)& NA\\
\textbf{$\gamma$ = 0.25 * $ \epsilon$} & 0.422 (1977/4681)& 0.375 (1409/3756)& 0.152 (28/184)& 0.079 (3/38)& NA\\
\textbf{$\gamma$ = 0.50 * $ \epsilon$} & 0.824 (3856/4681)& 0.802 (3011/3756)& 0.467 (86/184)& 0.316 (12/38)& NA\\
\textbf{$\gamma$ = 0.75  * $ \epsilon$} & 0.968 (4532/4681)& 0.968 (3634/3756)& 0.766 (141/184)& 0.684 (26/38)& 1.000 (1/1)\\
\textbf{$\gamma$ = $\epsilon$} & 1.000 (4681/4681)& 1.000 (3756/3756)& 1.000 (184/184)& 1.000 (38/38)& 1.000 (1/1)\\
\hline
\end{tabular}}
\qquad
\subfloat[Varying $\alpha$ - rate at which processes send messages]{\label{tab:fnvAlphaFiveProcInAppendix}\begin{tabular}{|l|l|l|l|}
\toprule
\textbf{Recall} & \multicolumn{1}{l|}{\textbf{$\alpha$ = 0.01}} & \multicolumn{1}{l|}{\textbf{$\alpha$ = 0.025}} & \multicolumn{1}{l|}{\textbf{$\alpha$ = 0.05}} \\
\midrule
\textbf{$\gamma$ = 0.10 * $ \epsilon$} & 0.007 (2/276)& 0.011 (2/174)& 0.022 (2/90)\\
\textbf{$\gamma$ = 0.15 * $ \epsilon$} & 0.011 (3/276)& 0.017 (3/174)& 0.022 (2/90)\\
\textbf{$\gamma$ = 0.20 * $ \epsilon$} & 0.011 (3/276)& 0.017 (3/174)& 0.022 (2/90)\\
\textbf{$\gamma$ = 0.25 * $ \epsilon$} & 0.022 (6/276)& 0.029 (5/174)& 0.044 (4/90)\\
\textbf{$\gamma$ = 0.50 * $ \epsilon$} & 0.123 (34/276)& 0.132 (23/174)& 0.178 (16/90)\\
\textbf{$\gamma$ = 0.75  * $ \epsilon$} & 0.409 (113/276)& 0.425 (74/174)& 0.544 (49/90)\\
\textbf{$\gamma$ = $\epsilon$} & 1.000 (276/276)& 1.000 (174/174)& 1.000 (90/90)\\
\hline
\end{tabular}}%
\caption{Percentage of Valid Snapshots detected during Conjunctive Predicate Detection using $\gamma$-extension out of all Valid Snapshots in the system.\label{tab:fnvConjPredInAppendix}}
\end{table*}%

\section{Precision and Recall results for application in \ref{sec:hlcgammaSMTExpeAnalysis}}

Tables \ref{tab:fpvTokenAppInAppendix} and \ref{tab:fnvTokenAppInAppendix} show the two-layered monitor's precision and recall values when deployed in the application presented in Section \ref{sec:hlcgammaSMTExpeAnalysis}. The corresponding time taken by the solver in each of these settings were presented in Figures \ref{fig:markwindgammadelta}, \ref{fig:markwindgammaeps}, \ref{fig:markwindgammaalpha}.

\begin{table*}
\centering{\textbf{No. of Valid Snapshots / Total Snapshots Detected  (Default values: n = 10, $\epsilon$ = 100, $\alpha$ = 0.1, $\delta$ = 10)}}\\
\subfloat[Varying $\delta$ - message delay]{\label{tab:fpvDeltaTokenInAppendix}\begin{tabular}{|l|l|l|l|l|l|l|}
\toprule
\textbf{Precision} & \multicolumn{1}{l|}{\textbf{$\delta$ = 1}} & \multicolumn{1}{l|}{\textbf{$\delta$ = 5}} & \multicolumn{1}{l|}{\textbf{$\delta$ = 10}} & \multicolumn{1}{l|}{\textbf{$\delta$ = 50}} & \multicolumn{1}{l|}{\textbf{$\delta$ = 100}} & \multicolumn{1}{l|}{\textbf{$\delta$ = 1000}} \\
\midrule
\textbf{$\gamma$ =  0.10 * $\epsilon$} & 1 (1/1)    & 1 (2/2)    & NA & 1 (1/1)    & 1 (1/1)    & 1 (1/1)\\
\textbf{$\gamma$ = 0.25 *  $\epsilon$} & 1  (4/4)   & 0.666667 (2/3)& 1  (3/3)   & 1   (2/2)  & 1  (2/2)   & 1 (3/3)\\
\textbf{$\gamma$ = 0.50 *  $\epsilon$} & 1 (5/5)    & 0.571429 (4/7) & 0.888889 (8/9) & 1  (6/6)   & 1   (4/4)  & 1 (5/5)\\
\textbf{$\gamma$ = 0.75 * $\epsilon$} & 0.5 (5/10) & 0.357143 (5/14)& 0.727273 (8/11)& 1  (10/10)   & 1 (8/8)    & 1 (8/8)\\
\textbf{$\gamma$ = $\epsilon$} & 0.333333 (5/15)& 0.263158 (5/19)& 0.571429 (8/14)& 1   (11/11)  & 1  (9/9)   & 1 (11/11)\\
\bottomrule
\end{tabular}}%
\qquad
\subfloat[Varying $\alpha$ - rate at which processes send messages]{\label{tab:fpvAlphaTokenInAppendix}\begin{tabular}{|l|l|l|l|l|l|}
\toprule
\multicolumn{1}{|l|}{\textbf{Precision}} & \multicolumn{1}{l|}{\textbf{$\alpha$ = 0}} & \multicolumn{1}{l|}{\textbf{$\alpha$ = 0.01}} & \multicolumn{1}{l|}{\textbf{$\alpha$ = 0.1}} & \multicolumn{1}{l|}{\textbf{$\alpha$ = 0.2}} & \multicolumn{1}{l|}{\textbf{$\alpha$ = 0.5}} \\
\midrule
\textbf{$\gamma$ = 0.10 * $\epsilon$} & 1    (1/1) & 1    (1/1) & NA & 1    (1/1) & 1 (3/3)\\
\textbf{$\gamma$ = 0.25 * $\epsilon$} & 1    (4/4) & 1    (3/3) & 1    (3/3) & 1    (1/1) & 1 (4/4)\\
\textbf{$\gamma$ = 0.50 * $\epsilon$} & 1    (6/6) & 1    (6/6) & 0.888889(8/9) & 1    (2/2) & 0.666667 (4/6)\\
\textbf{$\gamma$ = 0.75  * $\epsilon$} & 1    (7/7) & 1    (6/6) & 0.727273(8/11) & 0.4  (2/5) & 0.666667 (4/6)\\
\textbf{$\gamma$ = $\epsilon$} & 1    (10/10) & 0.888889(8/9) & 0.571429(8/14) & 0.222222(2/9) & 0.5 (4/8)\\
\bottomrule
\end{tabular}
}
\qquad
\subfloat[Varying $\epsilon$ - clock skew]{\label{tab:fpvEpsTokenInAppendix}\begin{tabular}{|l|l|l|l|l|l|}
\toprule
\textbf{Precision} & \multicolumn{1}{l|}{\textbf{$\epsilon$ = 1}} & \multicolumn{1}{l|}{\textbf{$\epsilon$ = 2}} & \multicolumn{1}{l|}{\textbf{$\epsilon$ = 5}} & \multicolumn{1}{l|}{\textbf{$\epsilon$ = 10}} & \multicolumn{1}{l|}{\textbf{$\epsilon$ = 100}} \\
\midrule
\textbf{$\gamma$ = 0.10 * $\epsilon$} & 1    (1/1) & NA & NA & 1    (2/2) & NA \\
\textbf{$\gamma$ = 0.25 * $\epsilon$} & 1    (1/1) & NA & 1    (5/5) & 1    (4/4) & 1 (3/3)\\
\textbf{$\gamma$ = 0.50 * $\epsilon$} & 1    (1/1) & 1    (3/3) & 1    (8/8) & 1    (8/8) & 0.888889 (8/9)\\
\textbf{$\gamma$ = 0.75  * $\epsilon$} & 1    (1/1) & 1    (3/3) & 1    (9/9) & 1    (11/11) & 0.727273 (8/11)\\
\textbf{$\gamma$ = $\epsilon$} & 1    (14/14) & 1    (3/3) & 1    (12/12) & 1    (11/11) & 0.571429 (8/14)\\
\bottomrule
\end{tabular}
}
\caption{Percentage of Valid Snapshots out of all snapshots detected during mutual exclusion detection using $\gamma$-extension.\label{tab:fpvTokenAppInAppendix}}
\end{table*}%
%
%
%
%
%
\begin{table*}
\centering{\textbf{No. of Valid Snapshots detected / No. of Valid Snapshots in the system (Default values: n = 10, $\epsilon$ = 100, $\alpha$ = 0.1, $\delta$ = 10)}}\\
\subfloat[Varying $\delta$ - message delay]{\label{tab:fnvDeltaTokenInAppendix}\begin{tabular}{|l|l|l|l|l|l|l|}
\midrule
\textbf{Recall} & \multicolumn{1}{l|}{\textbf{$ \delta$ = 1}} & \multicolumn{1}{l|}{\textbf{$ \delta$ = 5}} & \multicolumn{1}{l|}{\textbf{$ \delta$ = 10}} & \multicolumn{1}{l|}{\textbf{$ \delta$ = 50}} & \multicolumn{1}{l|}{\textbf{$ \delta$ = 100}} & \multicolumn{1}{l|}{\textbf{$ \delta$ = 1000}} \\
\midrule
\textbf{$ \gamma$ = 0.10* $\epsilon$} & 0.2 (1/5)   & 0.4 (2/5)  & 0 (0/8)    & 0.090909 (1/11) & 0.111111 (1/9) & 0.090909 (1/11)\\
\textbf{$ \gamma$ = 0.25* $\epsilon$} & 0.8 (4/5)  & 0.4 (2/5)  & 0.375 (3/8) & 0.181818 (2/11) & 0.222222 (2/9) & 0.272727 (3/11)\\
\textbf{$ \gamma$ = 0.50* $\epsilon$} & 1 (5/5)   & 0.8 (4/5)  & 1  (8/8)   & 0.545455 (6/11) & 0.444444 (4/9) & 0.454545 (5/11) \\
\textbf{$ \gamma$ = 0.75 * $\epsilon$} & 1    (5/5) & 1   (5/5)  & 1  (8/8)   & 0.909091 (10/11) & 0.888889 (8/9) & 0.727273 (8/11) \\
\textbf{$ \gamma$ = $\epsilon$} & 1 (5/5)    & 1  (5/5)   & 1  (8/8)   & 1   (11/11)  & 1  (9/9)   & 1 (11/11)\\
\bottomrule
\end{tabular}}
\qquad
\subfloat[Varying $\alpha$ - rate at which processes send messages]{\label{tab:fnvAlphaTokenInAppendix}\begin{tabular}{|l|l|l|l|l|l|}
\toprule
\multicolumn{1}{|l|}{\textbf{Recall}} & \multicolumn{1}{l|}{\textbf{$\alpha$ = 0}} & \multicolumn{1}{l|}{\textbf{$\alpha$ = 0.01}} & \multicolumn{1}{l|}{\textbf{$\alpha$ = 0.1}} & \multicolumn{1}{l|}{\textbf{$\alpha$ = 0.2}} & \multicolumn{1}{l|}{\textbf{$\alpha$ = 0.5}} \\
\midrule
\textbf{$\gamma$ = 0.10 * $\epsilon$} & 0.1  (1/10) & 0.125(1/8) & 0    (0/8) & 0.5  (1/2) & 0.75 (3/4)\\
\textbf{$\gamma$ = 0.25 * $\epsilon$} & 0.4  (4/10) & 0.375(3/8) & 0.375(3/8) & 0.5  (1/2) & 1 (4/4)\\
\textbf{$\gamma$ = 0.50 * $\epsilon$} & 0.6  (6/10) & 0.75 (6/8) & 1    (8/8) & 1    (2/2) & 1 (4/4)\\
\textbf{$\gamma$ = 0.75  * $\epsilon$} & 0.7  (7/10) & 0.75 (6/8) & 1    (8/8) & 1    (2/2) & 1 (4/4)\\
\textbf{$\gamma$ = $\epsilon$} & 1    (10/10) & 1    (8/8) & 1    (8/8) & 1    (2/2) & 1 (4/4)\\
\bottomrule
\end{tabular}%
}
\qquad
\subfloat[Varying $\epsilon$ - clock skew]{\label{tab:fnvEpsTokenInAppendix}\begin{tabular}{|l|l|l|l|l|l|}
\toprule
\textbf{Recall} & \multicolumn{1}{l|}{\textbf{$\epsilon$ = 1}} & \multicolumn{1}{l|}{\textbf{$\epsilon$ = 2}} & \multicolumn{1}{l|}{\textbf{$\epsilon$ = 5}} & \multicolumn{1}{l|}{\textbf{$\epsilon$ = 10}} & \multicolumn{1}{l|}{\textbf{$\epsilon$ = 100}} \\
\midrule
\textbf{$\gamma$ = 0.10 * $\epsilon$} & 0.071429(1/14) & 0    (0/3) & 0    (0/12) & 0.181818(2/11) & 0 (0/8)\\
\textbf{$\gamma$ = 0.25 * $\epsilon$} & 0.071429(1/14) & 0    (0/3) & 0.416667(5/12) & 0.363636(4/11) & 0.375 (3/8)\\
\textbf{$\gamma$ = 0.50 * $\epsilon$} & 0.071429(1/14) & 1    (3/3) & 0.666667(8/12) & 0.727273(8/11) & 1 (8/8)\\
\textbf{$\gamma$ = 0.75  * $\epsilon$} & 0.071429(1/14) & 1    (3/3) & 0.75 (9/12) & 1    (11/11) & 1 (8/8)\\
\textbf{$\gamma$ = $\epsilon$} & 1    (14/14) & 1    (3/3) & 1    (12/12) & 1    (11/11) & 1 (8/8)\\
\bottomrule
\end{tabular}
}
\caption{Percentage of Valid Snapshots detected during mutual exclusion detection using $\gamma$-extension out of all Valid Snapshots in the system.\label{tab:fnvTokenAppInAppendix}}
\end{table*}%

\end{document}